\documentclass[8.5pt,twocolumn]{revtex4-1}

\usepackage[version=3]{mhchem} 
\usepackage[T1]{fontenc}       
\usepackage{amsmath}	
\usepackage{setspace}	
\usepackage{graphicx}	
\usepackage{subfigure}	
\usepackage{xcolor}		
\usepackage{verbatim}	

\begin{document}





\title{Segregated structures of ring polymer melts near the surface : Molecular dynamics simulation study}
\vspace{0.6cm}

\author{Eunsang Lee}
\affiliation{Department of Chemistry, Seoul National University, Seoul 151-747, Korea}
\author{YounJoon Jung}
\affiliation{Department of Chemistry, Seoul National University, Seoul 151-747, Korea}
\email{ yjjung@snu.ac.kr}


\begin{abstract}
We study structural properties of a ring polymeric melt confined in a film in comparison to a linear counterpart using molecular dynamics simulations. Local structure orderings of ring and linear polymers in the vicinity of the surface are similar to each other because the length scale of surface-monomer excluded volume interaction is smaller than the size of an ideal blob of the ring. In a long length scale, while the Silberberg hypothesis can be used to provide a physical origin of confined linear polymer results, it no longer holds for a ring polymer case. We also present different structural properties of ring and linear polymers in a melt, including the size of polymers, an adsorbed amount, and the coordination number of a polymer. Our observation reveals that a confined ring in a melt adopts highly segregated conformation due to a topological excluded volume repulsion, which may provide a new perspective to understand the nature of biological processes, such as territorial segregation of chromosomes in eukaryotic nuclei.

\end{abstract}

\maketitle





\section{\label{sec-1} Introduction}

Understanding the nature of DNA packing during interphase in a nucleus for higher eukaryotes has been a long-lasting challenge in the field of biophysics. One of the most interesting features of DNA in a nucleus is chromosomal territories, in which chromatin fibers are not tangled to each other without making any knot, and each chromosome occupies a distinct territory.\cite{Cremer_NatRevGenet_2001,Bolzer_PLOSBiol_2005, Cremer_CurrOpinCellBiol_2006} Experiments have been performed to quantify the chromosome structure using, e.g., HiC for a contact probability \cite{Lieberman_Science_2009, Sexton_Cell_2012, Zhang_Cell_2012} and FISH for subchain sizes.\cite{Engh_Science_1992, Yokota_JCellBiol_1995, Langerak_PNAS_2009} In these works, the exponents of scaling relations for the contact probability $P_{\text c}(s)\sim s^{-\gamma}$ and for the subchain size $r(s)\sim s^{\nu}$ versus the genomic distance $s$ turned out to be $\gamma\approx 1$ and $\nu\approx 1/3$, respectively. These experimental data provide both intriguing and challenging questions to polymer physicists since DNA, being a polymer with open ends, seems to provide different scaling exponents from those of entangled linear polymers characterized by $\gamma\approx 1.5$ and $\nu\approx 1/2$.\cite{Rubinstein_PolPhys_2003, Doi_PolPhys_1986, Kremer_JChemPhys_1990} 

An early theoretical suggestion to interpret the DNA packing in terms of polymer physics involves an idea of a crumpled globule. \cite{Grosberg_JPhysFrance_1988, Grosberg_SoftMatter_2014} It is formed by a fast collapse of a single polymer as the solvent quality is quenched. Although the crumpled globule shares some characteristics with the chromosome, including knot-free configurations and the same fractal dimension, the idea still remains a hypothesis due to, e.g., long-term stability issue.\cite{Schram_JCP_2013} In recent years, inspired by aforementioned properties of crumpled globules, a ring in a melt has been proposed as a good candidate for elucidating the nature of DNA packing. \cite{Rosa_PLOSComputBiol_2014, Halverson_RepProgPhys_2014} Since both ends of the ring are connected to each other it possesses intrinsic topological constraints, non-knotting and non-concatenation, as represented in the chromosome structure.

Much attention has been paid to understand the effect of the topology of rings in a melt on their structures. In the early stage of these studies, it was understood that rings in a melt have an intermediate size characterized by the Flory exponent of $2/5$ in a long length scale, which was supported by Flory-type argument\cite{Cates_JPhys_1986} as well as simulation studies.\cite{Muller_PhysRevE_1996, Brown_JChemPhys_1998, Brown_PhysRevE_2001, Hur_Macromolecules_2006, Lee_MacromolRapidCommun_2015} Recently, however, numerical simulations of sufficiently long rings gave an evidence for an asymptotic behavior of the ring size with the Flory exponent of $1/3$ \cite{Muller_PhysRevE_2000, Halverson_JChemPhys_2011, Halverson_PhysRevLett_2012, Rosa_PhysRevLett_2014, Michieletto_SoftMatter_2014} as well as for the chromosome structures.\cite{Engh_Science_1992, Yokota_JCellBiol_1995, Langerak_PNAS_2009} Especially, it turns out that a universal behavior of the ring sizes was found, regardless of simulation models,\cite{Halverson_PhysRevLett_2012} which provoked many theoretical studies regarding a better understanding of the ring statics. Modified Flory argument implementing topological constraints of a ring also reproduced the asymptotic behavior of a ring size and crossover behavior between an ideal and a compact polymer statistics.\cite{Sakaue_PhysRevLett_2011, Sakaue_PhysRevE_2012} In these works, topological constraints are dealt with the effective excluded volume interactions. It is also fascinating that a simulation snapshot of ring polymers in a melt recently reported in Ref.\citenum{Vettorel_PhysToday_2009} allows one to recall the image of segregated chromosomes in a nucleus.\cite{Cremer_NatRevGenet_2001} Motivated by the above works, showing knot-free, unentangled conformations similar to the interphase chromosome structures, study of the physics of rings in a melt has recently come into the spotlight.

Coming back to the problem of the DNA packing, it may be useful to review recent research activities focusing on the question why the chromatin fiber does not show reptation behavior even though it has free ends to reptate. One of the studies reveals that the reptation time of chromatin fiber is much longer than a cellular lifetime because its length is much larger than the nuclear size.\cite{Rosa_PLOSComputBiol_2014} Bulky ends formed by repeated telomeric DNA sequence\cite{Olovnikov_JTheorBiol_1973, Greider_Nature_1989} and some parts of chromatin fiber attached to inner surface of the nuclear envelope are also suspected to prevent the chromosome from relaxing to the intermixed state through reptation.\cite{Halverson_RepProgPhys_2014} Along the line of these studies, it is worthwhile to investigate the adsorption of the DNA on the nuclear envelope in the language of polymer physics. 

A lot is known for the structures of confined linear polymer in a melt via experiments,\cite{Wilson_ApplClaySci_1999, Tsukruk_Langmuir_2001, Smith_PhysRevLett_2003} computer simulations,\cite{Daoulas_Macromolecules_2005, Virgiliis_EurPhysJE_2012, Pandey_JChemPhys_2012, Smith_PhysRevLett_2003} and theories.\cite{Roe_JChemPhys_1974, Scheutjens_JPhysChem_1979, Scheutjens_JPhysChem_1980,  Eisenriegler_JChemPhys_1982} It is also interesting to note that two-dimensional linear polymers in a melt, i.e., the limit of the ultrathin film, show highly segregated conformations and amoeba-like dynamics,\cite{Carmesin_JPhysFrance_1990, Meyer_EurPhysJ_2007, Wittmer_PhysRevLett_2010, Meyer_JChemPhys_2010} which resemble the physics of rings in a melt.\cite{Halverson_JChemPhys_2011_2}  Compared to the linear case, a relatively little attention has been paid to the confined ring systems for the needs of understanding confined circular biopolymers, e.g., circular DNAs. Starting from the self-consistent field theory of a single adsorbed ring polymer on the surface,\cite{Lent_Macromolecules_1987} some experiments\cite{Persson_NanoLett_2009, Witz_PhysRevLett_2011, Reisner_PhysRevLett_2005}, simulations\cite{Witz_PhysRevLett_2011, Benkova_Macromolecules_2012, Sheng_PhysRevE_2012, Sakaue_EurophysLett_2010} and a theory\cite{Sakaue_EurophysLett_2010} of confined ring polymers in a dilute regime have been performed. However, the confinement and the surfaces effects on the concentrated rings have remained a challenge. Problems in synthesizing the non-concatenated, non-knotted and monodisperse ring polymers make one difficult to investigate physics of confined rings in a melt by an experimental approach.\cite{Bras_SoftMatter_2014}

In this study, therefore, employing molecular dynamics simulations, we study perturbation of confined ring structures by a surface and show how the surface affects the static properties of rings, e.g., ordering of local structures, sizes of rings, adsorbed amounts, and the coordination numbers of a polymer. The aim of this work is not only to investigate the ring structures when the surface effect is added to intrinsic topological constraints, but to look for new perspective for the nature of DNA packing in a nucleus.

The rest of this paper is organized as follows. We first explain a flexible Kremer-Grest bead-spring model and a film system designed by implicit soft walls in Sec. \ref{sec-2}. In Sec. \ref{sec-3}, after observing well-defined layering structures of monomers and bonds near the surface, we present structure perturbation on the scale of the polymer size. The static properties altered by the surface interaction and the topological repulsion are also provided. Concluding remarks follow in the final section.

\section{\label{sec-2} Simulation Methods}

\begin{figure}[t]
	\centering
	\includegraphics[width=2.5in]{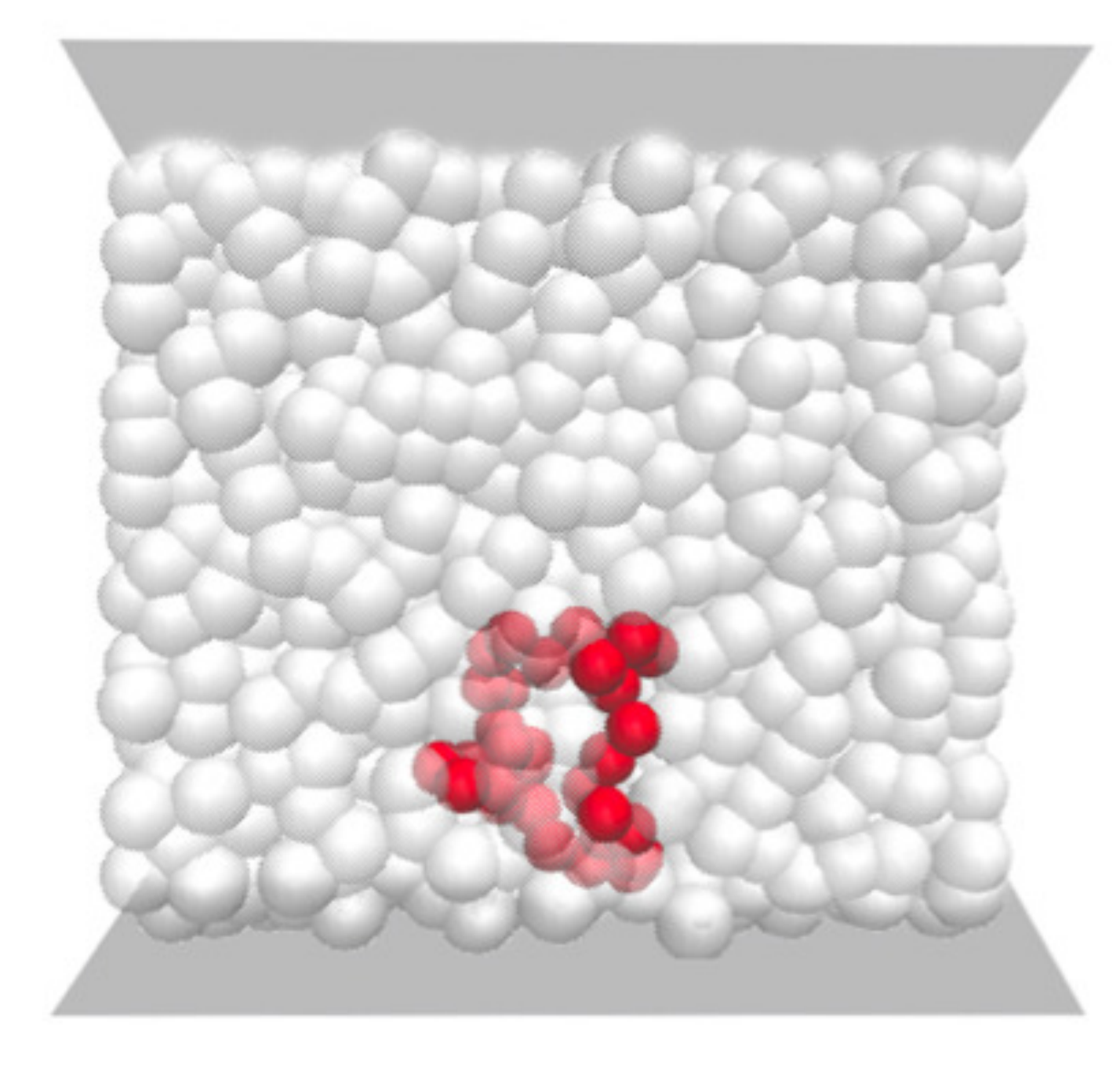}
	\caption{\label{fig-system} Schematic description of a simulation system of $N=32$ ring polymers. A ring polymer colored by red represents one of adsorbed polymers.}
\end{figure}

In order to study static properties of linear and ring polymers near the surface in a thin film morphology, we performed molecular dynamics simulations of polymers via the flexible Kremer-Grest model (Fig. \ref{fig-system}).\cite{Kremer_JChemPhys_1990} In this model, a polymer consists of beads having unit mass $m$. Non-bonded beads interact with each other through repulsive Lennard-Jones (Weeks-Chandler-Anderson, WCA) potential (Eqn. \ref{eqn-wca}) with the unit energy $\epsilon$ and the unit length $\sigma$. The covalent bond between beads is described by finite extensible nonlinear elastic (FENE) potential (Eqn. \ref{eqn-fene}) to prevent bond crossing. No angle potential is included to describe flexible polymers.

\begin{equation}
\label{eqn-wca}
U_{\textrm{pair}}(r_{ij})=\left \{ \begin{array}{ll}
4 \epsilon [ ( {\sigma}/{r_{ij}} ) ^{12}- ( {\sigma}/{r_{ij}} ) ^{6} + \frac{1}{4} ], &
r_{ij} < 2^{1/6} \sigma \\
0, & \textrm{otherwise}, 
\end{array} \right.
\end{equation}

\begin{equation}
\label{eqn-fene}
U_{\textrm{bond}}(r_{ij})=\left \{ \begin{array}{ll}
-0.5kR_{0}^{2} \ln [ 1-( {r_{ij}}/{R_{0}} ) ^{2} ], & r_{ij}<R_{0}, \\
\infty, & \textrm{otherwise}
\end{array} \right.
\end{equation}
In Eqn. \ref{eqn-fene}, $k=30\epsilon/\sigma^{2}$ and $R_{0}=1.5\sigma$ are used to suppress the bond length fluctuation. Implicit soft walls perpendicular to $z$-axis are located at the bottom and top of $z$-axis. Two different non-bonded interaction potentials between the surface and a monomer are used to compare chain properties between attractive and repulsive surfaces. The attractive surface is represented by Lennard-Jones potential using the same parameters with WCA potential described above, except for the cutoff distance $2.5\sigma$ and attraction strength $3\epsilon$. The repulsive surface is modeled by WCA potential with the same parameter in the monomer-monomer repulsion. The unit time is scaled by $\tau=\sigma \sqrt{m/\epsilon}$. We employ the weak temperature coupling method, Langevin thermostat (Eqn. \ref{eqn-langevin}) in which a temperature, an integration time step, a friction coefficient are set to be $1.0\epsilon/k_{\text{B}}$, $0.01\tau$ and $0.5\tau^{-1}$, respectively. 
\begin{equation}
\label{eqn-langevin}
\ddot{\mathbf{r}_{i}}=\nabla \sum_{i\neq j} U_{ij}-\Gamma\dot{\mathbf{r}_{i}}+\mathbf{W}_{i}(t)
\end{equation}
All simulations are conducted by LAMMPS package.\cite{Plimpton_JCompP_1995}

\begin{table}[t]
\caption{\label{tab-system}System description: $N$, $M$, $L_{z}$, $L_{x}=L_{y}$, and $t_{\text{eq}}$ are, respectively, the degree of polymerization, the number of polymers in a system, a film separation, a system size along $x$ and $y$ direction, and an equilibration time. The monomer densities of all systems are $0.85\sigma^{-3}$.}
\setlength{\tabcolsep}{10pt}
\centering
\begin{tabular}{rrrrr}
\hline\hline
$N$ & $M$ & $L_{z}^{'}$ & $L_{x}=L_{y}$ & $t_{\text{eq}}$  \\
\hline
32 & 128 & 20.00 & 15.98 & $1.0\times10^{5}$ \\ 
64 & 128 & 20.00 & 22.60 & $1.0\times10^{5}$ \\ 
128 & 128 & 40.00 & 31.49 & $1.0\times10^{5}$ \\ 
256 & 256 & 40.00 & 44.53 & $5.0\times10^{5}$ \\
512 & 256 & 60.00 & 51.18 & $2.0\times 10^{7}$ \\
\hline\hline
\end{tabular}
\end{table}

\begin{figure}[t]
	\centering
	 \includegraphics[width=2.5in]{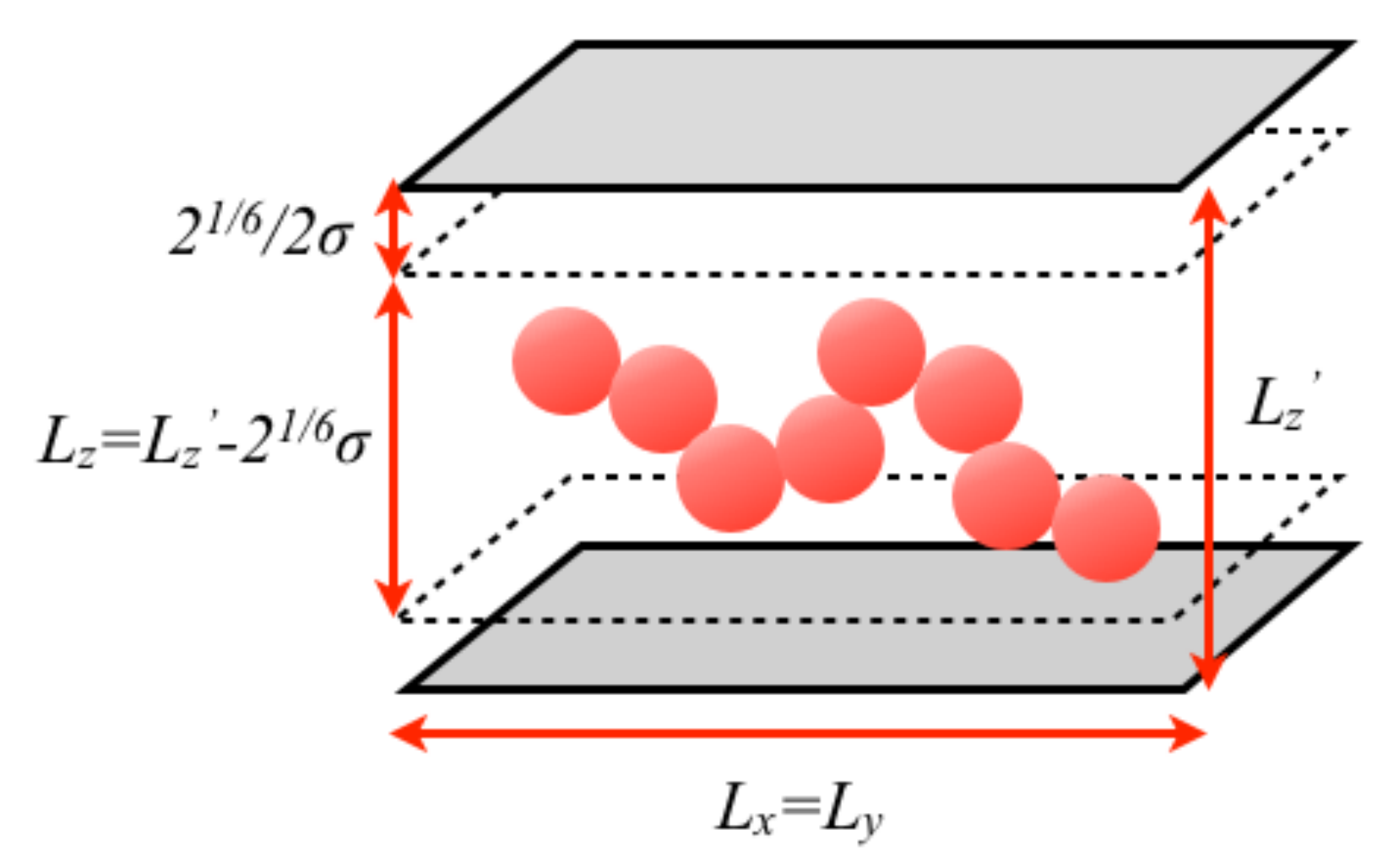}
\caption{\label{fig-film} Schematic description of the surface-monomer excluded volume. It should be taken into account when one calculates the system volume.}
\end{figure}

We simulate monodisperse polymer systems of five different lengths, $N=$ 32, 64, 128, 256, and 512 for both linear and ring polymers (Table \ref{tab-system}). The number density of monomers is $0.85\sigma^{-3}$, which is a typical monomer density of a melt for the bead-spring model. To take into account the excluded volume of surfaces, we calculate the system volume by  $V=L_{x}L_{y}L_{z}=L_{x}L_{y}(L_{z}^{'}-2^{1/6}\sigma)$, where $L_{z}^{'}$ is the actual separation between two surfaces (Fig.  \ref{fig-film}). To avoid a bridged polymer, the separation between two surfaces should be larger than $4R_{\text{g,bulk}}$, where $R_{\text{g,bulk}}$ is the radius of gyration in the bulk phase. The bridging structure is not observed at all in our simulation trajectories. To minimize the finite size effect, we enlarge the system size along $x-$ and $y-$ directions and also increase the number of polymers in the system as $N$ increases. Obviously, periodic boundary condition is used only in $x-$ and $y-$ directions (Table \ref{tab-system}).

To focus on the properties of non-concatenated, non-knotted ring polymers we adopt the following preparation scheme. At first, after fixing the film separation to the desired value, we place monomers with perfect circular morphology and the planes of the circles are perpendicular to $z-$ axis. Centers of circles are placed on square lattice sites in the $xy$-plane. All monomers are located at the center between two surfaces. To avoid concatenation, the lattice constant of the square lattice should be larger than a diameter of a circle. To do so, we set the lattice constant to $2.5r=2.5N2^{1/6}\sigma/2\pi$, because $N2^{1/6}\sigma\approx2\pi r$ for $N>>1$, where $r$ is a radius of a circle. At high pressure of $P=5.0\epsilon/k_{\text{B}}$ in $x$- and $y$- directions, short NPT simulations are performed until desired values of $L_{x}$ and $L_{y}$ which satisfy $\rho=NM/L_{z}L_{x}L_{y}$ are obtained. In these preparation procedures, attractive interactions between chains are turned on by changing the cutoff distance of WCA potential to $r_{c}=2.5\sigma$. For all systems, desired densities are attained within $5\times10^{3}\tau$ (Table \ref{tab-system}).

In order to equilibrate systems, NVT simulations are performed starting from the initial configuration obtained by the above method. We again change the cutoff distance of WCA potential from $r_{c}=2.5\sigma$ to $r_{c}=2^{1/6}\sigma$, to turn off intermonomer attractions. Equilibrations are executed during time larger than five times of Rouse time of polymers in the bulk phase. The lengths of production runs of all systems are $10^{7}\tau$ and the configurations are saved in every $10^{2}\tau$ to get ensemble averages.

\section{ \label{sec-3} Results and Discussions}

\subsection{Chain Non-ideality at the Vicinity of the Surface}

\begin{figure}[t]
	\centering
	\includegraphics[width=3.1in]{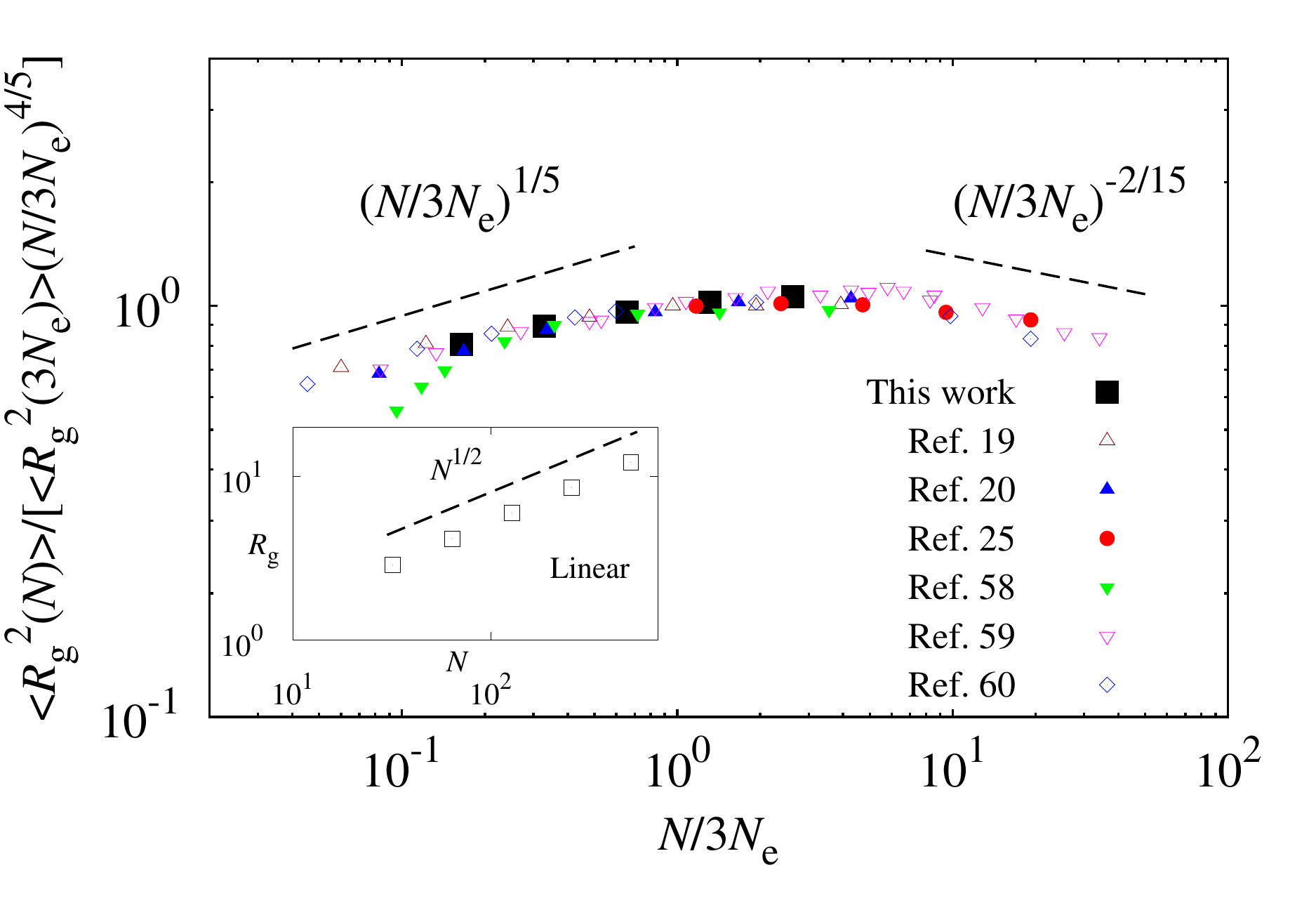}
\caption{\label{fig-bulkrg} Rescaled gyration radii multiplied by $N^{4/5}$ of rings as a function of degree of polymerization normalized by entanglement length, $N_{e}$, which was originally provided in the work of Ref. \citenum{Halverson_PhysRevLett_2012}. Those obtained in our simulations (black filled squares, $N_{\text{e}}=70$) and other previous works in Ref. \citenum{Muller_PhysRevE_1996} ($N_{\text{e}}=85$), Ref. \citenum{Brown_JChemPhys_1998} ($N_{\text{e}}=40$), Ref. \citenum{Halverson_JChemPhys_2011} ($N_{\text{e}}=28$), Ref. \citenum{Hur_Macromolecules_2011} ($N_{\text{e}}=140$), Ref. \citenum{Suzuki_JChemPhys_2009} ($N_{\text{e}}=40$), and Ref. \citenum{Vettorel_PhysBiol_2009} ($N_{\text{e}}=175$) are shown by different point types. Inset shows the gyration radii of our linear polymers, which are exactly scaled by $N^{1/2}$.}
\end{figure}

Before studying properties of confined polymers, we first calculate the sizes of linear and ring polymers in the bulk phase as shown in Table \ref{tab-bulksize}. The density of monomers in the bulk is $0.85\sigma^{-3}$ as well as in the film system. As expected, the fully flexible linear chains used in our simulation show ideal statistics characterized by the fractal dimension 2 (inset of Fig. \ref{fig-bulkrg}). However, physics of ring polymers varies with the degree of polymerization $N$. Halverson et. al.\cite{Halverson_PhysRevLett_2012} found striking merging behavior of the ring sizes for many different models, which provides a clear crossover from Gaussian regime represented by $R_{\text g}\sim N^{1/2}$ to compact globular regime by $R_{\text g}\sim N^{1/3}$ as shown in Fig. \ref{fig-bulkrg}. Rings in this study correspond to $0.4N_{\text{e}}$ through $8N_{\text{e}}$ ($N_{\text e}\approx 70$ for the flexible bead-spring model \cite{Everaers_Science_2004}) and long rings of $N=256$ and 512 lie on the intermediate regime in which the ring size is approximately proportional to $N^{2/5}$. They are expected to show the different physical properties from confined linear polymers due to the different Flory exponents.
\begin{table}[t]
\caption{\label{tab-bulksize} Gyration radii and end-to-end distances (spanning distances) for linear and ring polymers in the bulk phase.}
\setlength{\tabcolsep}{10pt}
\centering
\begin{tabular}{rrrrr}
\hline\hline
 & \multicolumn{2}{c}{Linear} & \multicolumn{2}{c}{Ring} \\
\cline{2-5}
$N$ & $R_{\text{g}}^2$ & $R_{\text{e}}^2$ & $R_{\text{g}}^2$ & $R_{\text{e}}^2$ \\ 
\hline
32 & 8.21 & 49.55 & 4.41 & 14.06 \\
64 & 17.32 & 104.56 & 8.51 & 26.14 \\
128 & 35.85 & 215.80 & 16.02 & 48.00 \\
256 & 72.76 & 436.89 & 29.51 & 86.64 \\
512 & 149.13 & 896.49 & 52.92 & 152.74 \\
\hline\hline
\end{tabular}
\end{table}

\begin{figure}[t]
	\centering
		\includegraphics[width=2.8in]{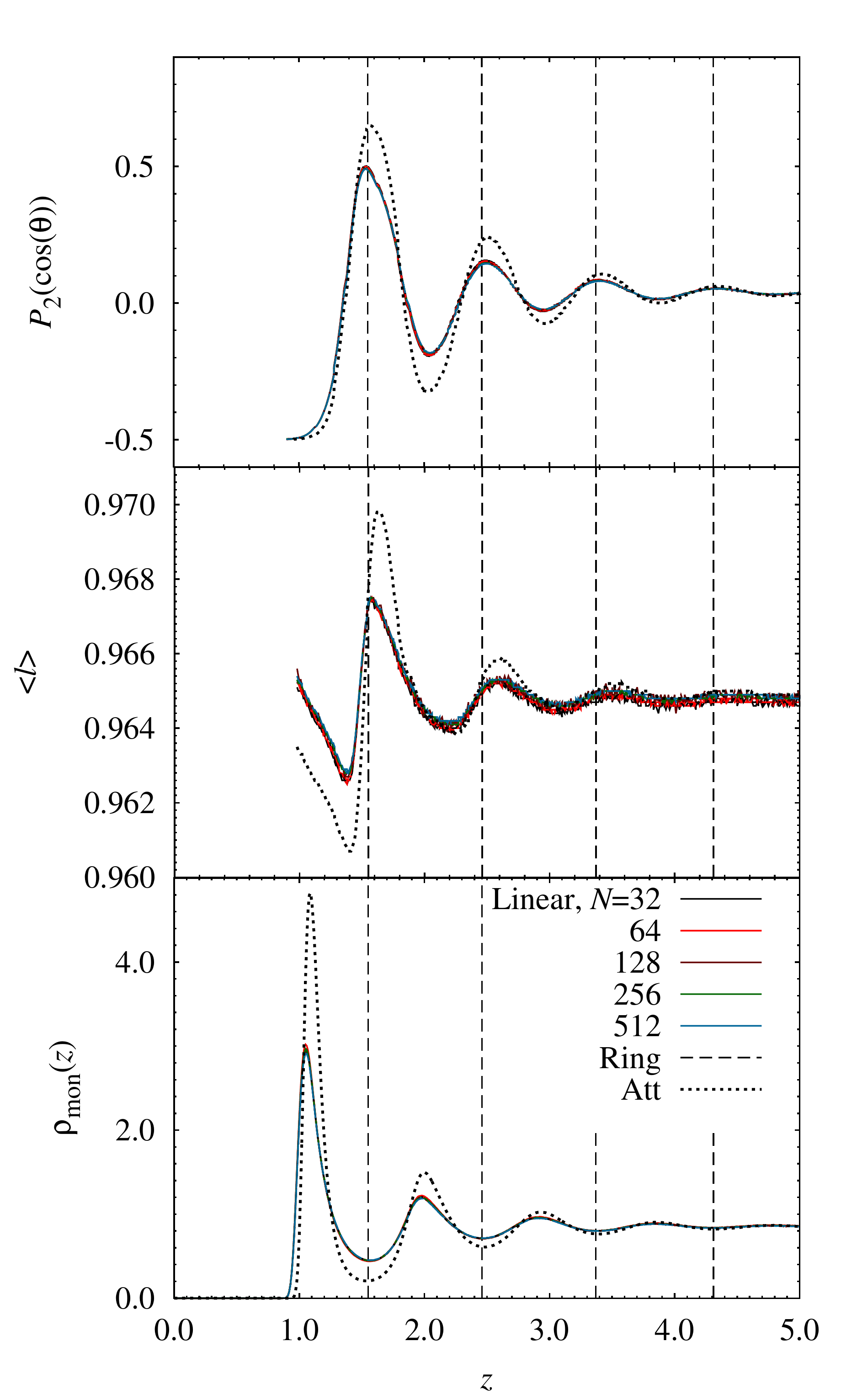}
\caption{\label{fig-layer} Variations of monomer densities (bottom), bond lengths (middle), and bond orientations (top) as a function of distance between the monomer and the surface. Black dashed lines represent the position at which densities of monomers are minimal. In all figures, results of attractive surfaces for linear polymers of $N=512$ are represented by black dotted line.}
\end{figure}

Returning our attention to the confined polymers, it would be instructive to study the effect of local non-ideality in the vicinity of the surface induced by the excluded volume interaction between a monomer and a surface. We display variations of monomer densities, bond lengths, and bond alignments for linear and ring polymers as a function of the distance of a monomer from the surface $z$ in Fig. \ref{fig-layer}. In this figure, it is shown that the oscillation of monomer densities fades away beyond $z\approx 4\sigma$ by forming four layers regardless of $N$. Results of $N=512$ linear polymers within attractive surfaces are also plotted by black dotted lines in the same graphs. Linear and ring polymers with other $N$ give almost the same variations with the $N=512$ linear polymer (not provided here). The attractive surface leads to the polymer structure with more order, but the length scale hardly depends on the attraction strength. In the absence of a strong and long-range attraction such as Coulombic potential, the length scale of solely the excluded volume interaction between the monomer and the surface, usually $\approx \sigma$, is shorter than the polymer size. The surface contributes to forming the first layer whose thickness is about $\sigma$. The ordered monomers in the first layer help the monomers to be ordered in the second layer, and in turn, the third and fourth layers are also formed in a layer-by-layer fashion. During this procedure, ordering becomes weak as $z$ increases. Such a layering structure is observed not only in polymer systems but in the various confined liquid systems, e.g., ionic liquids in the graphine double layers.\cite{DeYoung_JPhysChemC_2014}

Bond lengths and orientations are also oscillate near the surface. Second Legendre polynomial, $P_{2}(\cos(\theta))$, is calculated to obtain bond alignments by such that, 
\begin{equation}
\label{eqn-secondLP}
P_{2}(\cos(\theta))=\frac{3}{2}\langle\cos^{2}(\theta)\rangle-\frac{1}{2}\text{,}
\end{equation}
where $\theta$ is an angle between the bond vector and the normal to the surface. This value varies from -0.5 meaning that the bonds are parallel, to 1.0 being perpendicular to the surface. When bonds align randomly, it is averaged out to be zero. Not only the monomer densities, bond lengths and alignments oscillate with the same amplitude and the wavelength regardless of $N$. In Fig. \ref{fig-layer}, black dashed vertical lines indicate the distances from the surface at which the monomer density is minimal. In the region where the monomers are sparsely distributed, the bonds are elongated and align perpendicular to the surface. In the dense region, in contrast, they contract and align with the surface. Note that only the magnitude of oscillation varies as the interaction strength between the surface and the monomer changes. These observations indicate that the bond length interaction hardly affects the structure ordering, and the oscillations of bond properties are chiefly determined by the monomer positions.

In addition to independence of the chain length and the surface attraction on the local structure ordering, the response of the confined polymer structure is insensitive to the chain topology (dashed lines in Fig. \ref{fig-layer}). The surface perturbs the polymer structures in a layer-by-layer fashion and the size of ring polymers is typically larger than the length scale of the surface-monomer excluded volume interaction. Thereby, the surface can not recognize whether the interacting monomers belong to the linear or the ring polymer. While an elongated layering structure is expected in the system of higher monomer density, all systems produce the same length scale of this non-ideal structure in our simulation condition regardless of the chain length, the strength of surface interaction and the chain topology. For the further discussion, it is useful to keep in mind that the length scale of the chain non-ideality, $\approx 4\sigma$, is even smaller than the sizes of $N=128<3N_{\text e}$ ring polymers giving ideal statistics characterized by the Flory exponent $\nu=1/2$ (see Table \ref{tab-bulksize} and Fig. \ref{fig-bulkrg}).

\subsection{Segment Properties}

\begin{figure}[t]
	\centering
	\includegraphics[width=2.2in]{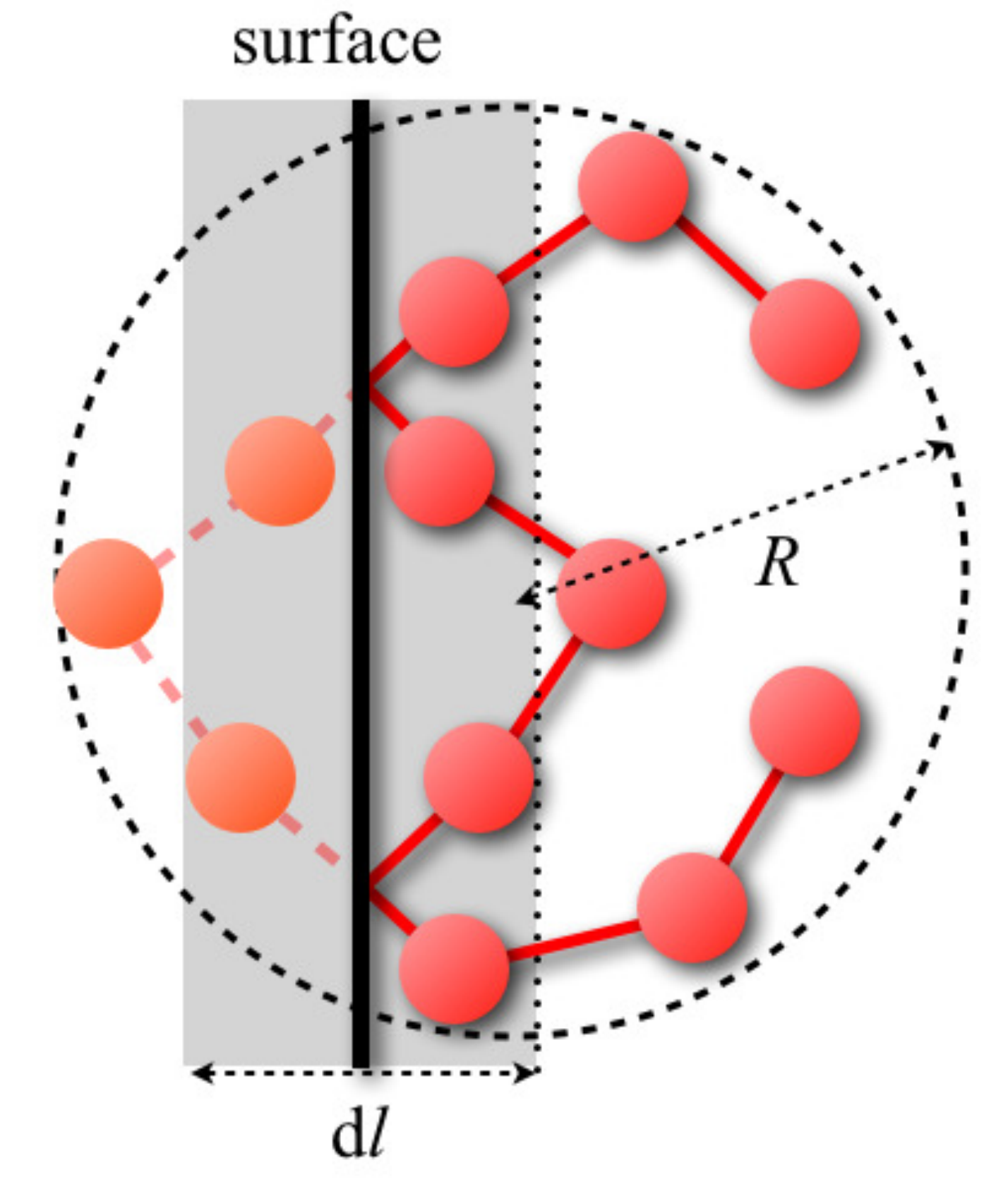}
\caption{\label{fig-surfpol} Schematic description (side view) of the conformation transfer through a reflecting boundary. Shaded monomers with orange color behind the surface are reflected in front of the surface. $R$ is the polymer size if it were in the bulk phase and $\text{d}l$ is the height of the disk represented by black shaded area.}
\end{figure}

To study the effect of the surface-monomer interaction on the polymer structure, it is instructive to take into account the Silberberg hypothesis \cite{Silberberg_JCollInterSci_1982} based on the random walk with the reflective surface, which explains the physics of the confined linear polymer melts (Fig. \ref{fig-surfpol}). In this argument, assuming a reflecting boundary in the middle of an ideal chain, one can imagine that monomers which would have been located behind the boundary in the bulk phase are reflected through the boundary. Because of no excluded volume interaction between monomers in an ideal chain, the reflected monomers do not perturb the structure of remaining part of the chain, the consequence of which is that the parallel size of the chain does not vary from the bulk one. 

\begin{figure}[t]
	\begin{center}
	\includegraphics[width=2.8in]{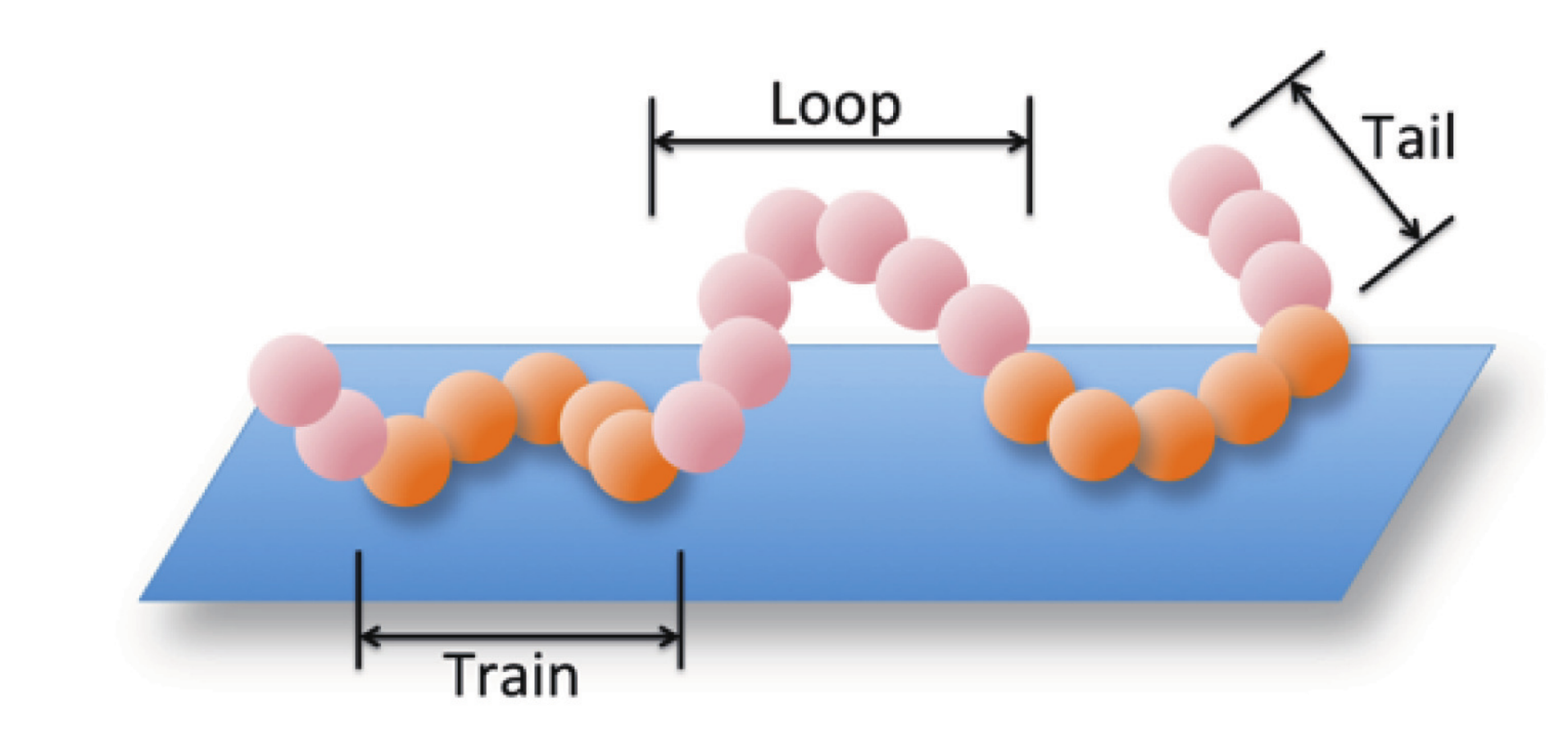}
	\end{center}
\caption{\label{fig-segdefin} Schematic descriptions of a loop, a train, and a tail in a linear polymer. Monomers colored with orange are adsorbed on the surface. Definitions of segments of a ring polymer are the same as those of a linear polymer.}
\end{figure}
\begin{figure}[t]
	\centering
	\subfigure[]{\label{fig-segnum}\includegraphics[width=3.1in]{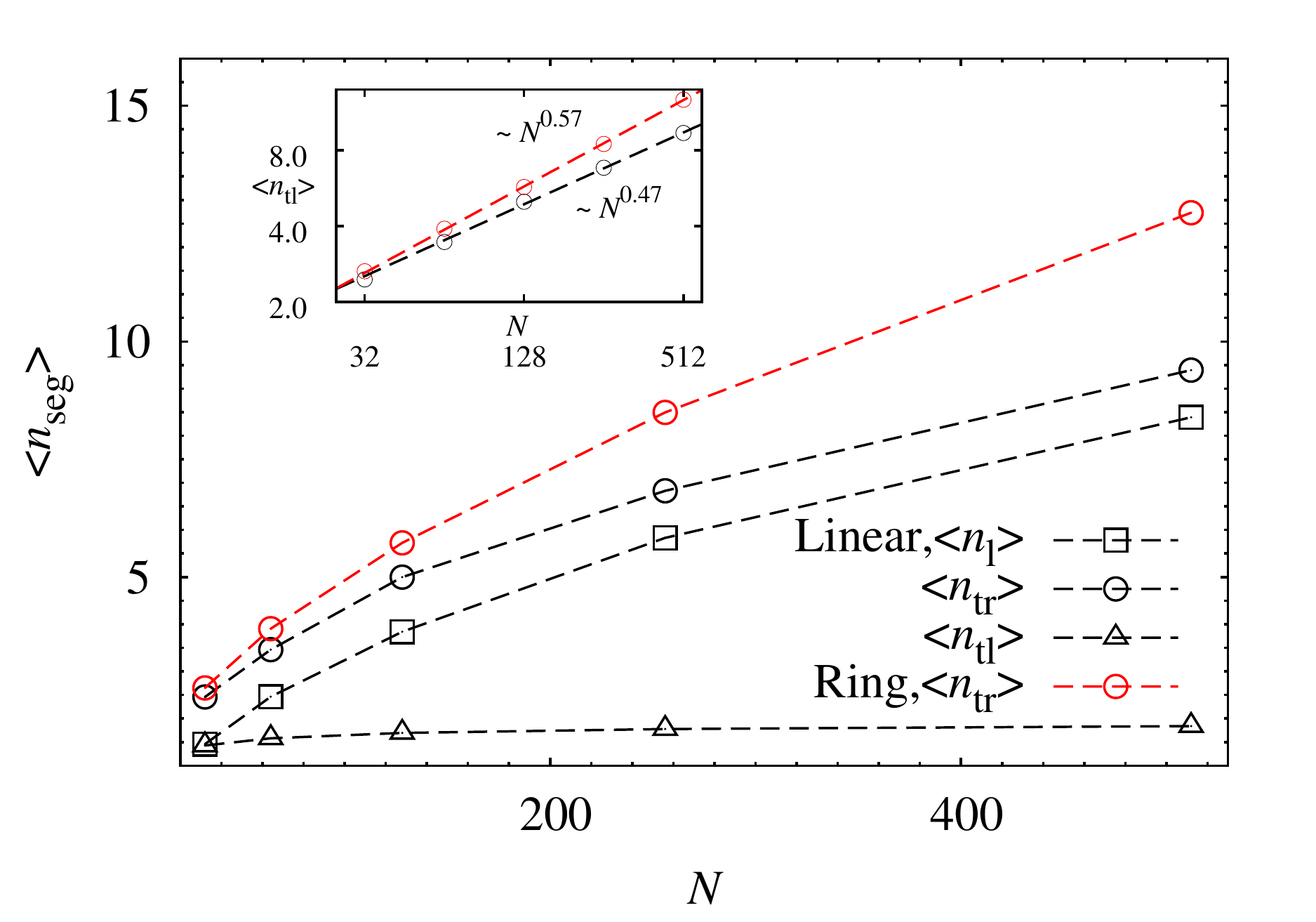}}
	\subfigure[]{\label{fig-segsize}\includegraphics[width=3.1in]{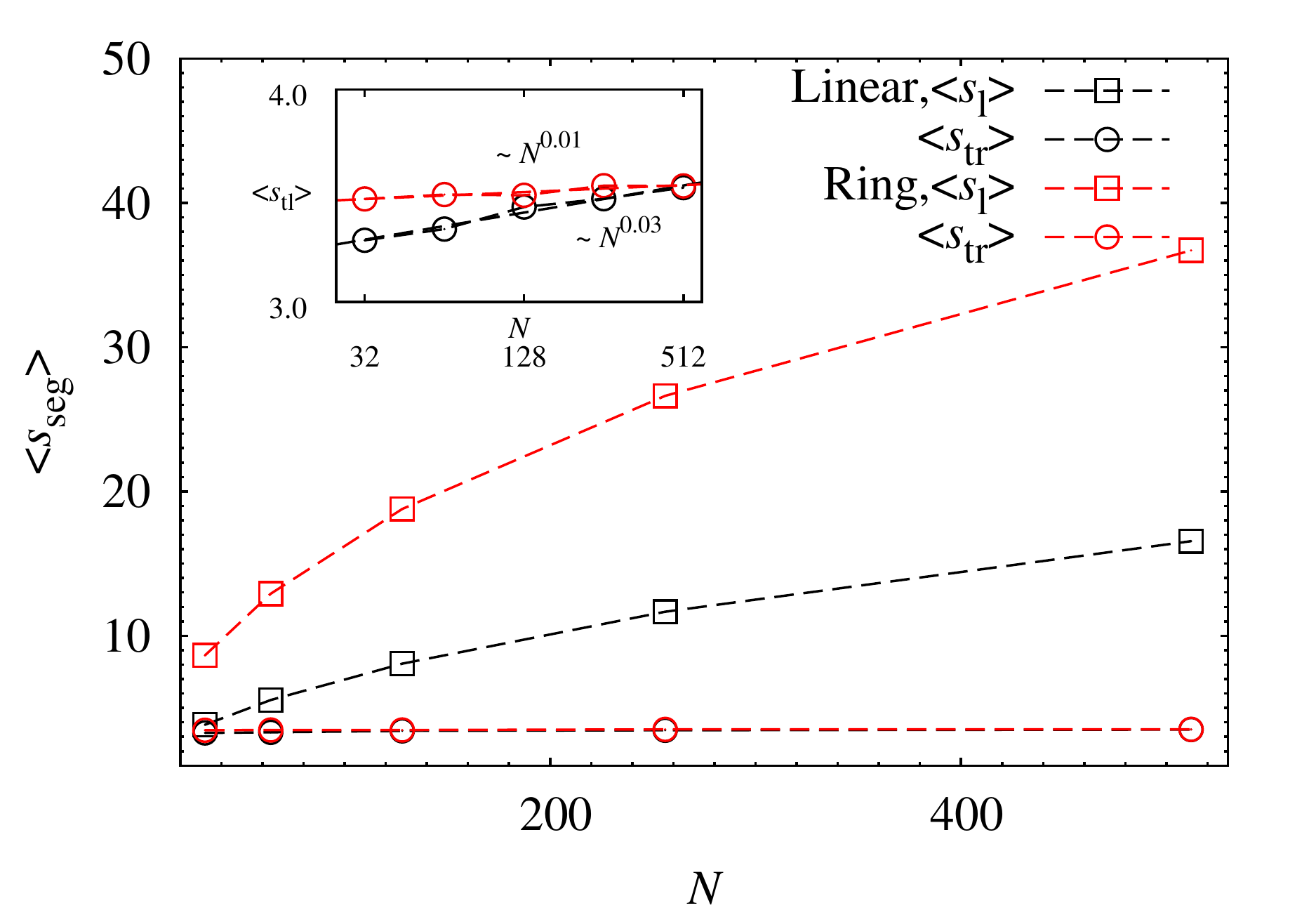}}
\caption{\label{fig-segment} (a)The average number and (b)the average size of loops (squares), trains (circles), and tails (triangles) for linear (black) and ring (red) polymers. The sizes of trains for linear polymers overlap with those for rings in (b). Insets in both figures show log-log plots of the number and the size of trains. Both average values can be scaled by $N^{\gamma}$ and the fitted exponents are represented in each figure.}
\end{figure}

The monomer excluded volume in the real system, however, cannot be screened out as argued in the structure ordering near the surface. Here we provide segment properties of adsorbed polymers, such as the sizes and the numbers of trains, loops and tails to understand the interplay between the chain non-ideality in the vicinity of the surface and the overall structure perturbation. The definitions of segments are the same as in previous works.\cite{Scheutjens_JPhysChem_1980, Virgiliis_EurPhysJE_2012} As depicted in Fig. \ref{fig-segdefin}, successive adsorbed monomers are considered to a train, and the segment of successive non-adsorbed monomers whose both ends are adsorbed on the surface is a loop. A tail is defined by consecutive non-adsorbed monomers in both ends of a linear chain. Obviously, a ring polymer does not have a tail. From now on, we discuss only the results of systems with the repulsive surface since the surface interaction strentgh modeled by short-range Lennard-Jones potential does not change structural properties of polymers significantly. Figure \ref{fig-segnum} presents the average numbers of loops $\langle n_{\text{l}} \rangle$, trains $\langle n_{\text{tr}}\rangle$, and tails $\langle n_{\text{tl}}\rangle$ per an adsorbed chain as a function of $N$, where the adsorbed chain is defined such that at least one monomer belonging to it is located in the first layer represented in Fig. \ref{fig-layer}, which is identical to the definition in the Scheutjens-Fleer theory.\cite{Scheutjens_JPhysChem_1979, Scheutjens_JPhysChem_1980} In Fig. \ref{fig-system}, the highlighted ring polymer is an example of an adsorbed chain. According to the definitions, the number of trains of a ring polymer is the same as that of loops if the perfectly adsorbed polymer is absent. The relations, $n_{\text{tr}}=n_{\text{l}}+1$ and $0\leq n_{\text{tl}}\leq 2$, also hold for linear polymers. In accordance with the recent work of Ref. \citenum{Virgiliis_EurPhysJE_2012}, $\langle n_{\text{tr}}\rangle$ and $\langle n_{\text{l}} \rangle $ of linear polymers simultaneously grow with $N$, but $\langle n_{\text{tl}} \rangle$ very slowly increases below 2. $\langle n_{\text{tr}} \rangle $ and $\langle n_{\text{l}}\rangle$ of ring polymers also increase with $N$, and are larger than those of linear ones. Figure \ref{fig-segsize} shows the average sizes of loops $\langle s_{\text l}\rangle$, trains $\langle s_{\text {tr}} \rangle$, and tails $\langle s_{\text {tl}}\rangle $ per each segment. We find that $\langle n_{\text{tr}}\rangle $ and $\langle s_{\text{tr}}\rangle$ follow power laws as $N$ varies with the scaling exponents, $\gamma_{\text{n}}$ and $\gamma_{\text{s}}$, respectively, such that $\langle n_{\text{tr}}\rangle \sim N^{\gamma_{\text{n}}}$ and $\langle s_{\text{tr}}\rangle \sim N^{\gamma_{\text{s}}}$ (Insets of Fig. \ref{fig-segnum} and \ref{fig-segsize}). The fitted exponents for linear and ring polymers are listed in Table \ref{tab-adsexp}. It is intriguing to note that the sizes of trains for both linear and ring polymers, which slowly vary with $N$, are very similar to each other. The sizes of trains are determined only by the very local structure of polymers close to the surface. We argue that the length scale of structure ordering due to the surface is smaller than the that of chain ideality even of the ring polymers in a bulk phase. This indistinguishability between linear and ring polymer in a very short length scale leads to the same train size.

\begin{table}[t]
\caption{\label{tab-adsexp} Fitted scaling exponents for an adsorbed amount, $\delta$, the average number of trains, $\gamma_{\text{n}}$, and the average size of trains, $\gamma_{\text{s}}$, versus $N$ in the film and the bulk systems.}
\setlength{\tabcolsep}{10pt}
\centering
\begin{tabular}{rrrrrr}
\hline\hline
 & \multicolumn{2}{c}{Film} & &\multicolumn{2}{c}{Bulk} \\
\cline{2-3}\cline{5-6}
& Linear & Ring & & Linear & Ring\\
\hline
$\delta$ & 0.50 & 0.42 & & 0.49 & 0.42 \\
$\gamma_{\text{n}}$ & 0.47 & 0.57 & & 0.48 & 0.58 \\
$\gamma_{\text{s}}$ & 0.03 & 0.01 & & 0.03 & 0.00 \\
\hline\hline
\end{tabular}
\end{table}

\begin{figure}[t!]
	\centering
	\subfigure[]{\label{fig-trprob}\includegraphics[width=3.0in]{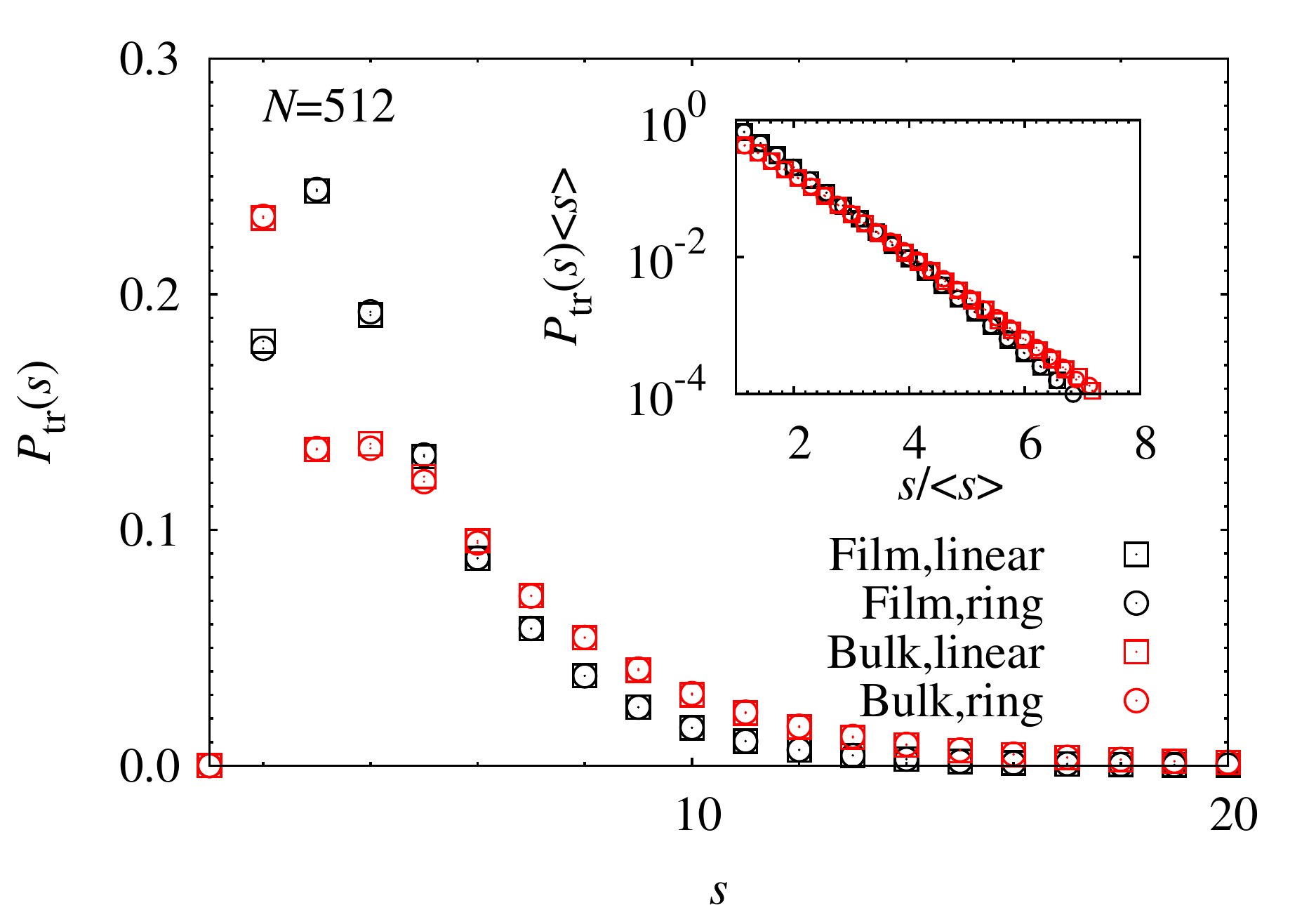}}
	\subfigure[]{\label{fig-lprob}\includegraphics[width=3.0in]{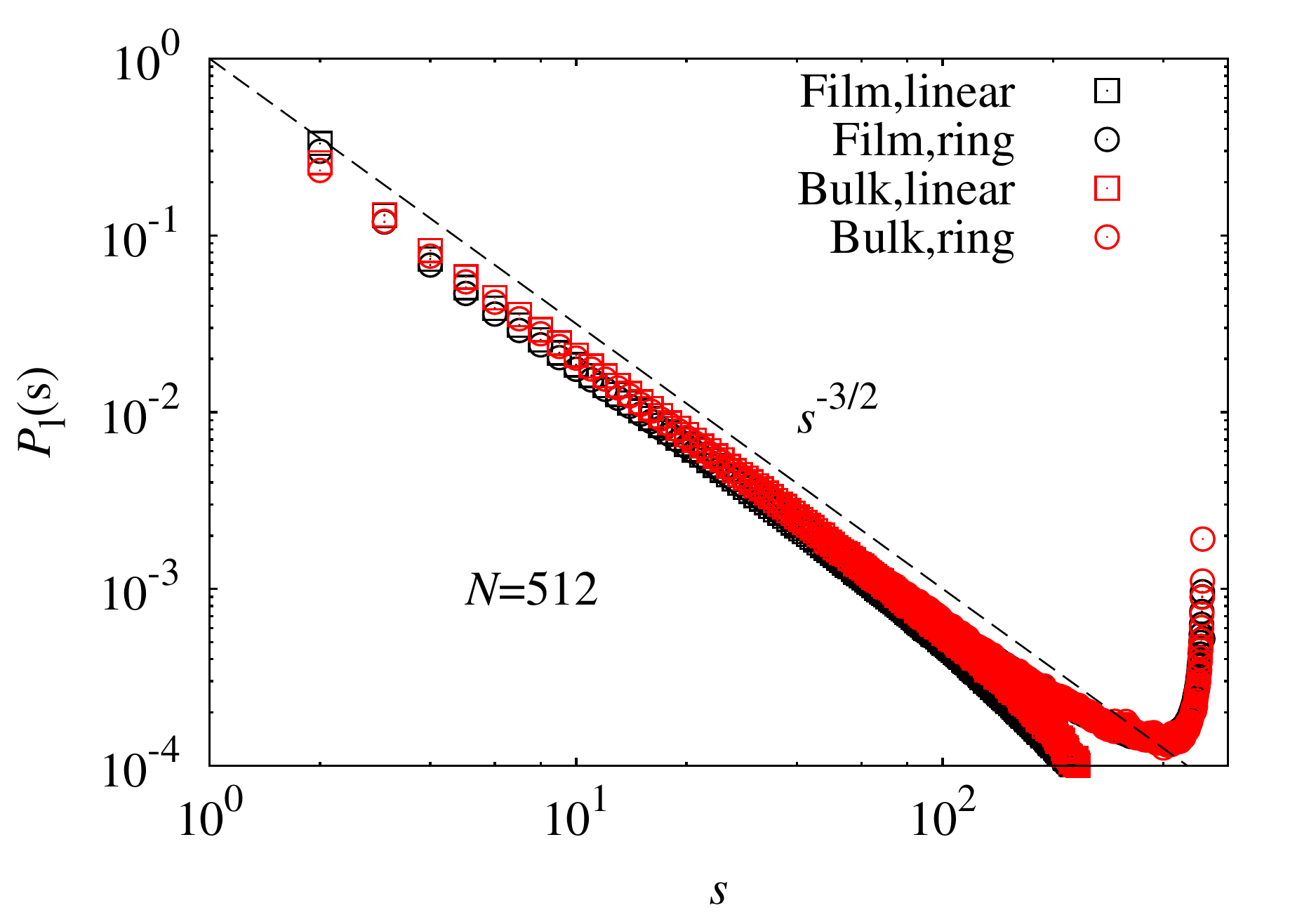}}
\caption{\label{fig-segprob} The probability distributions of (a) train and (b) loop sizes for linear (squares) and ring (circles) polymers of $N=512$. An inset of (a) is a log-linear plot of the train size distribution which shows exponential decay of probability distributions. Symbols colored by red represent results obtained in bulk system. In (b), all symbols merge into a single line which is scaled by $P_{\text l}(s)\sim s^{-3/2}$ except for $s>N/2$ loops for rings.}
\end{figure}

The probability distributions of segment sizes also support the above argument of the local indistinguishability between polymer topologies. Figure \ref{fig-segprob} displays the size distributions of (a) trains and (b) loops for $N=512$ linear and ring polymers. As expected from the previous analytical\cite{Scheutjens_JPhysChem_1980, Hoeve_JChemPhys_1965, Eisenriegler_JChemPhys_1982} and numerical\cite{Virgiliis_EurPhysJE_2012, Sarabadani_JChemPhys_2014} studies for the ideal chains, the train size distribution of linear polymers follows an exponential form, $P(s_{\text{tr}})/\langle s_{\text{tr}}\rangle \approx\exp({-s_{\text{tr}}/\langle s_{\text{tr}}\rangle})$, and the loop size distribution decays as a power law, $P(s_{\text{l}})\sim s^{-3/2}$, where $s$ is a contour length. Notice that the train size distributions of linear and ring polymers are the same with each other, which means that the train size of the ring polymer melt can be explained in terms of ideal chain statistics. The fact that loop size distribution of the ring polymer is also similar with linear ones in small $s$ also supports the argument of indistinguishability. Apparently, a high looping probability in large $s$ of the ring polymer is originated from the topology of connected chain ends.

To explain the difference in the average numbers of segments between linear and ring polymers, we evaluate the segment properties of the bulk polymers in the following procedure. Based on the idea of random walk with the reflective boundary (Fig. \ref{fig-surfpol}), we place a fictitious slab parallel to the xy-plane and in the middle of $z-$axis of the simulation box whose thickness is two times of the layer thickness in Fig. \ref{fig-layer}, such that $\text{d}l=2\times1.05\sigma=2.10\sigma$. We define a train in this system as successive monomers located in this slab, and a loop as successive monomers not located in the slab in analogy to the definition in the film system. The size distributions of trains and loops of bulk polymers well reproduce the results of confined polymers in Fig. \ref{fig-segprob}. The fact that fitted exponents of the relations $\langle n_{\text{tr}}\rangle \sim N^{\gamma_{\text n}}$ and $\langle s_{\text{tr}}\rangle\sim N^{\gamma_{\text s}}$ listed in Table \ref{tab-adsexp} are also in agreement with those of confined polymers reveals the local ideality of a ring. Using the Silberberg's idea, the number of segments that visit the slab is regarded as the number of trains, which is closely related to the overall structure of polymer, especially, to the self-monomer density. The self-monomer density of a ring polymer in a melt is higher than that of a linear polymer.\cite{Halverson_JChemPhys_2011} Thereby the segments of the ring visit the slab frequently, resulting in the higher average number of trains.

\subsection{Center of Mass Density}

\begin{figure}[t]
	\centering
	\includegraphics[width=3.1in]{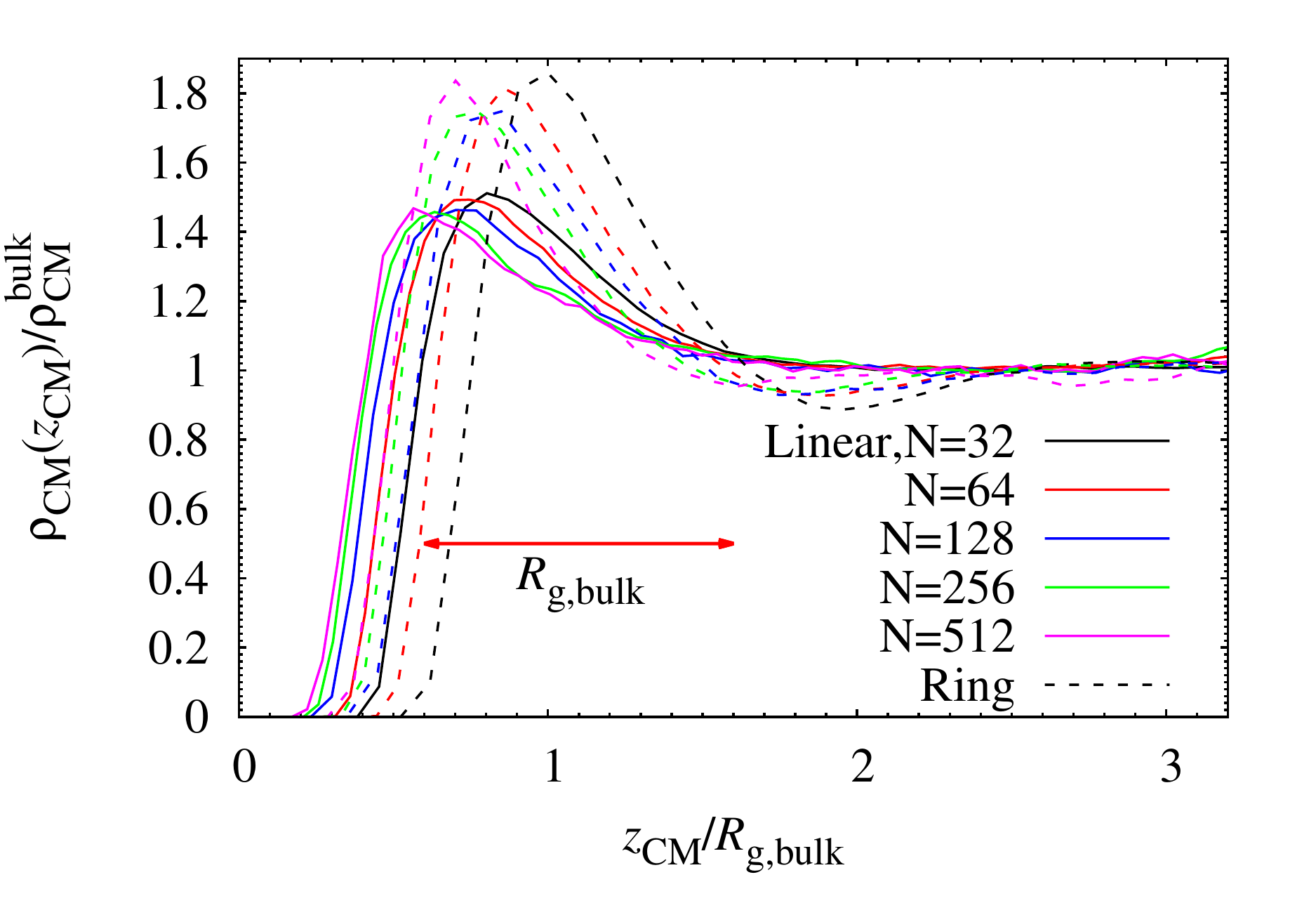}
\caption{\label{fig-comz} Densities of polymer center-of-mass normalized by the bulk density as a function of the distance from the surface also normalized by a radius of gyration of a bulk polymer for linear (solid) and ring (dashed lines) polymers. A red arrow indicates the $R_{\text {g,bulk}}$.}
\end{figure}

Up to now, we have focused on the local structure near the surface. The observation of the average numbers of trains motivates us to study the overall structure perturbation on the scale comparable to the polymer size. To do so, we first provide the density of polymer center-of-mass as a function of a distance of a polymer center-of-mass from the surface, $z_{\text{CM}}$, also normalized by the bulk radius of gyration, $R_{\text {g,bulk}}$ (Fig. \ref{fig-comz}). Packing structure of polymer molecules toward the surface is observed as well as the monomer ordering. It is obvious that the size of polymers at $z_{\text{CM}}<R_{\text{g,bulk}}$ is influenced by the surface as represented by a red arrow in Fig. \ref{fig-comz}. One can find a noticeable difference between linear and ring polymers. Rings are more densly packed toward the surface than linear polymers reflected by the high peaks of the center-of-mass density profiles at $z\approx R_{\text{g,bulk}}$. These high densities are compensated by the small densities at around $z\approx2R_{\text{g,bulk}}$. This trend resembles the packing of monomers which indicates that, unlike the linear polymers, the ring acts like a globular object rather than the ideal chain. It can be also understood by the topological excluded volume interaction imposed by the non-concatenation and the non-knotting constraints.\cite{Sakaue_PhysRevLett_2011, Sakaue_PhysRevE_2012} Another observation of non-monotonous peak heights of rings at $z\approx R_{\text{g,bulk}}$ with increasing $N$ also supports this argument, which can be understood by two contributions of the  local non-ideality and the topological repulsion. For small $N$, due to the comparable size of a ring to the length scale of the monomer packing structure, rings are densely packed. As $N$ increases, the surface-induced non-ideal structure becomes a small part of a ring, which weakens ordering of the molecules. If $N$ continuously grows to the asymptotic regime ($\nu=1/3$), the strong topological repulsion makes the ring again ordered.

\subsection{Polymer Size Near the Surface}

\begin{figure}[t]
	\centering
	\includegraphics[width=3.1in]{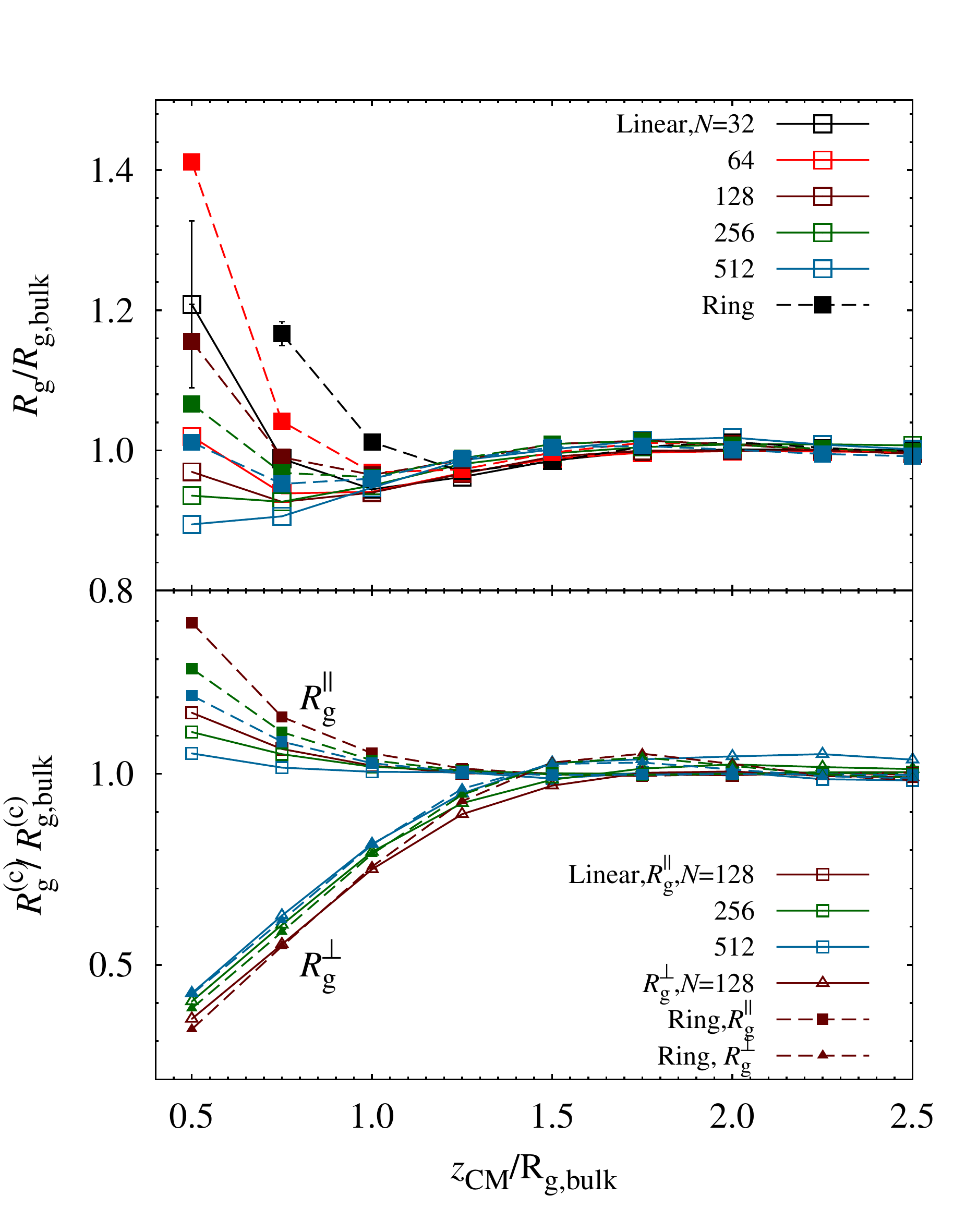}
\caption{\label{fig-rgslab} Normalized gyration radii (top) and their parallel and perpendicular components to the surface (bottom) as a function of the distance of a molecule center-of-mass from the surface, $z_{\text{CM}}$, for linear (open symbols) and ring (filled symbols) polymers. In the bottom, squares and triangles represent parallel and perpendicular sizes to the surface, respectively. In the upper figure, error bars for both polymers of $N=32$ are shown. Errors in $z_{\text{CM}}/R_{\text{g,bulk}}\geq0.75$ are smaller than the size of symbols.}
\end{figure}

Motivated by the topological interaction of a ring on the large scale, we now study the non-ideality of a ring polymer on the scale comparable to its size. According to the above argument, a long ring of $N>3N_{\text e}$ in a bulk is composed of ideal blobs of the size $\zeta\sim g^{1/2}$, where $g$ is the number of monomers in this blob, expected to $g < 3N_{\text e}$. These blobs are densely packed with each other in space characterized by $R\approx \zeta (N/g)^{\nu}\sim N^{\nu}$, where $\nu=2/5$ for long rings in this study. It is worthwhile to apply the ring to the argument of the conformation transfer through the reflective boundary. Reflected blobs in front of the boundary experience topological interactions imposed by the non-knotting constraint, which makes the ring expand. In Fig. \ref{fig-rgslab}, we plot the gyration radii normalized by $R_{\text{g,bulk}}$ versus $z_{\text{CM}}/R_{\text{g,bulk}}$. As the molecule approaches to the surface, it slightly shrinks at $z\approx R_{\text{g,bulk}}$, but again swells in the vicinity of the surface for both linear and ring polymers. The increased amount near the surface is greater for ring polymers than that for linear counterparts. It is clear to see the parallel, $R_{\text{g}}^{||}$, and perpendicular components, $R_{\text{g}}^{\perp}$, of gyration radii to the surface as shown at the bottom of Fig. \ref{fig-rgslab}. $R_{\text g}^{\perp}$'s for all systems merge into a single line within errorbars, which is in agreement with the above argument that the sizes of the polymers in $z_{\text {CM}}<R_{\text{g,bulk}}$ are affected. The validity of Silberberg hypothesis is underpinned by the obeservation that $R_{\text g}^{||}$'s of the linear systems are not much deviated from their bulk sizes, in agreement with the previous work.\cite{Sarabadani_JChemPhys_2014} For this linear case, a slight expansion of chain sizes in the vicinity of the surface is due to the non-ideality by the surface-monomer excluded volume interaction whose scale is up to $4\sigma$. For the ring polymer case, however, not only this short-range surface interaction but the topological repulsion enlarge the ring size along the parallel direction to the surface.

\begin{figure}[t]
	\centering
		\subfigure[]{\label{fig-adslin}\includegraphics[width=1.5in]{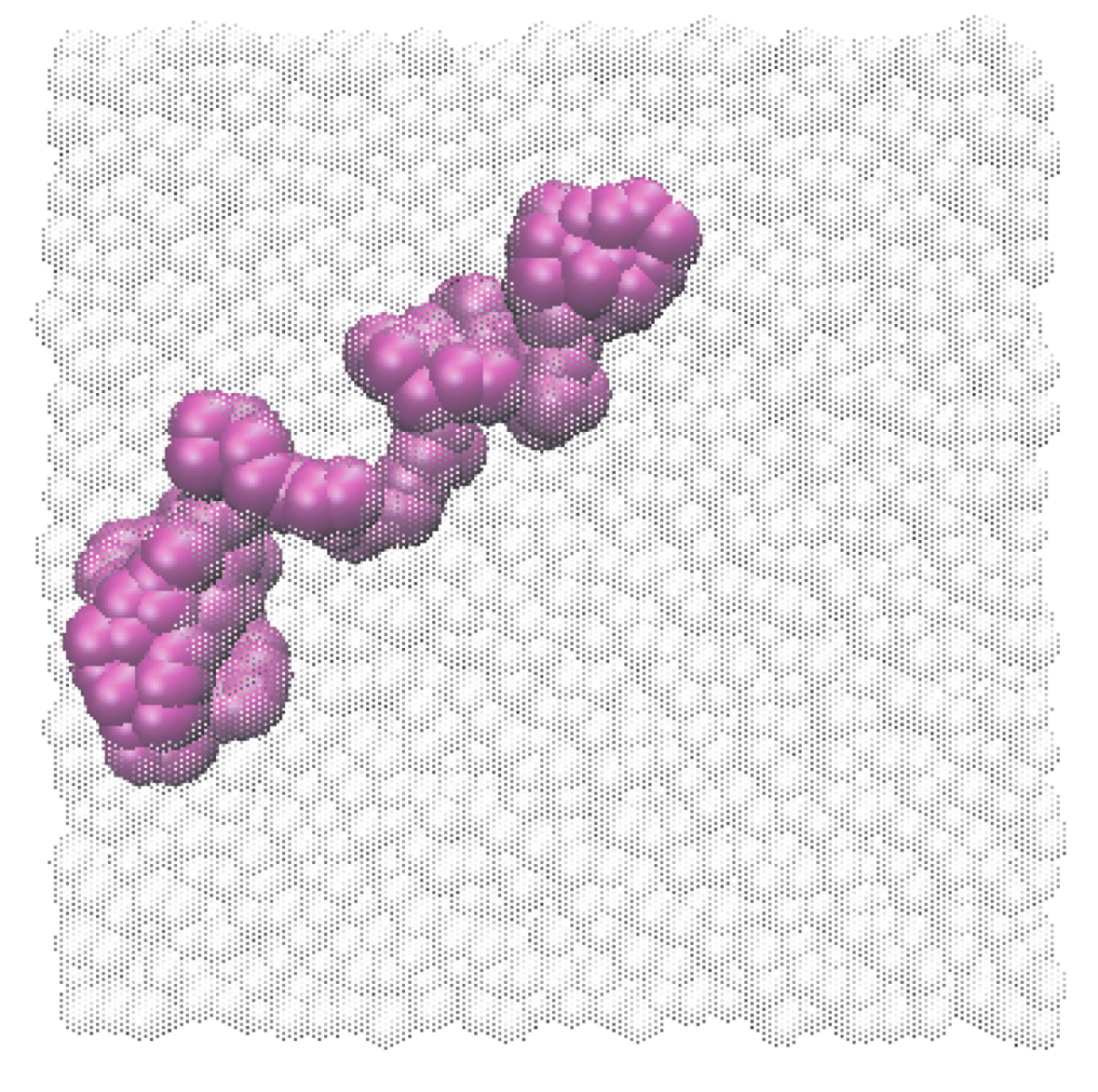}}
		\subfigure[]{\label{fig-adscir}\includegraphics[width=1.5in]{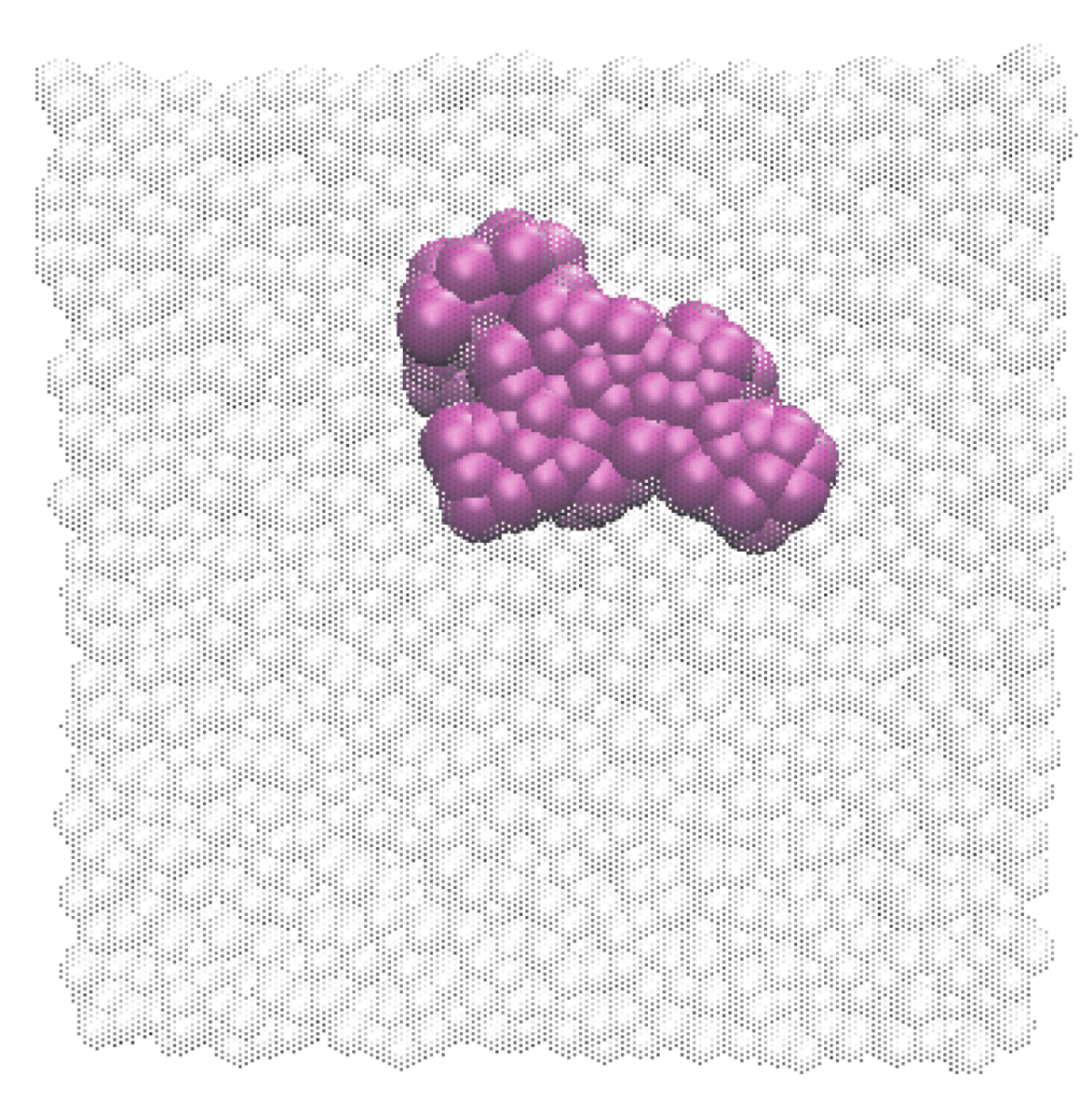}}
\caption{\label{fig-adspolsnap} Top-view snapshots of $N=128$ (a) linear and (b) ring polymers which are adsorbed on the surface. The surface interacting with highlighted polymers is located in front the page.}
\end{figure}

Figure \ref{fig-adspolsnap} represents top-view snapshots of our simulations of $N=128$ linear and ring polymers adsorbed on the surface. This figure clearly shows the difference of self-monomer densities of adsorbed polymers around their center-of-mass between linear and ring polymers. Linear polymers can penetrate into the space occupied by the other adsorbed linear polymer highlighted by pink color in Fig. \ref{fig-adslin} even at close to the surface. In contrast, a ring has hardly ever allows other rings to be tangled. Because of the topological excluded volume interaction, blobs in a ring repel each other, which causes the swelling of the ring under confinement.

\subsection{Adsorbed Amount}

\begin{figure}[t]
	\centering
	\includegraphics[width=3.1in]{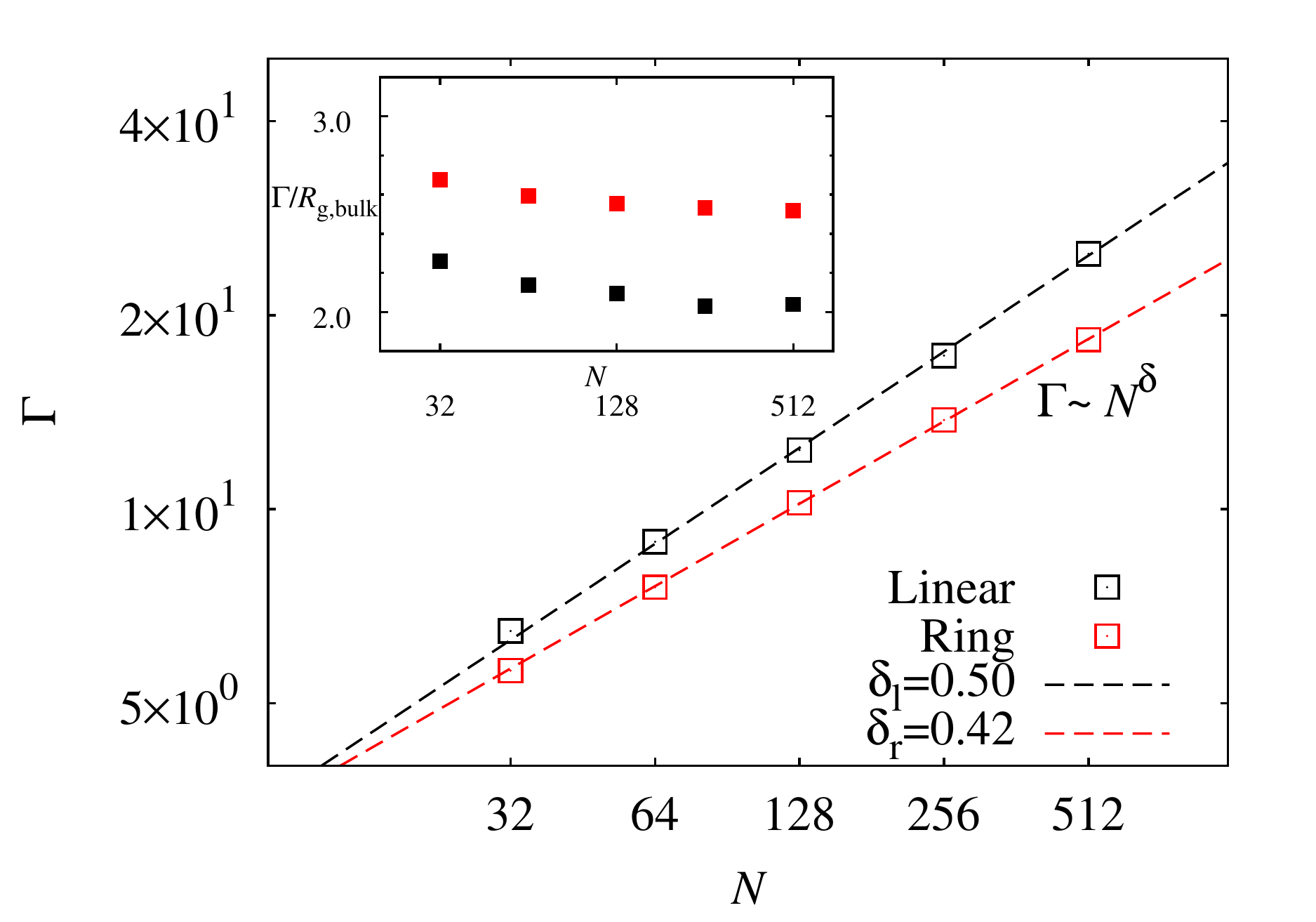}
\caption{\label{fig-adsamt} Log-log plots of adsorbed amounts, $\Gamma's$, as a function of $N$ for linear (black) and ring (red) polymers. In both cases, adsorbed amounts are proportional to the $N^{\delta}$ and fitted exponents are expressed in legends. Inset shows the adsorbed amounts divided by gyration radii in the bulk, $\Gamma/R_{\text{g,bulk}}$, versus $N$.}
\end{figure}

We now present another physical property of confined polymers, an adsorbed amount $\Gamma$, whose behavior depends on the chain ideality and the topological interactions. We define $\Gamma$ as the total mass of adsorbed chains per unit area. Figure \ref{fig-adsamt} shows $\Gamma$ as a function of $N$ for both polymers. The adsorbed amount of a linear system increases with $N$ with the power law, $\Gamma\sim N^{\delta}$, with the exponent $\delta_{\text l}=0.5$, which is well consistent with several previous studies. \cite{Scheutjens_JPhysChem_1979, Scheutjens_JPhysChem_1980, Virgiliis_EurPhysJE_2012} One intriguing observation is a behavior of the adsorbed amount for the ring system. It also follows a power law, but the exponent is somewhat smaller than that of the linear polymer, $\delta_{\text r}=0.42$. Interestingly, these two exponents of linear and ring polymers seem to be close to their Flory exponents, $\nu_{\text{l}}$ and $\nu_{\text{r}}$, respectively. As pointed out earlier, a linear chain in a melt follows ideal statistics with $\nu_{\text{l}}=1/2$, but the sizes of $N=256$ and 512 ring polymers lie on the intermediate regime with $\nu_{\text{r}}\approx2/5$ between an ideal chain and a compact globule (Fig. \ref{fig-bulkrg}). In order to compare two exponents, $\nu$ and $\delta$, we also plot the ratio of the adsorbed amount to the size, $\Gamma / R_{\text{g,bulk}}$ as a function of $N$ in the inset of Fig. \ref{fig-adsamt}. In this figure, the fact that the ratio is constant in asymptotic limit of $N\to\infty$ means that $\nu\approx\delta$. Using the picture of conformation transfer (Fig. \ref{fig-surfpol}), we can prove this relation for the linear polymer case via mean-field approach. The number of adsorbed monomers per a single adsorbed polymer, $n_{\text{ads}}$, can be evaluated by counting the number of monomers transpiercing the disk of the height $\text{d}l$ and the radius $R$, which is described by shaded area in Fig. \ref{fig-surfpol}. Because the self monomer density, $\rho_{\text{self}}$, is proportional to $N/R^{3}$ and the volume of the disk is $V_{\text{disk}}\sim R^{2}\text{d}l$, the number of adsorbed monomers per an adsorbed polymer is calculated as:
\begin{equation}
\label{eqn-idealads}
n_{\text{ads}}\approx V_{\text{disk}}\times\rho_{\text{self}}\approx R^{2}\text{d}l\times \frac{N}{R^{3}} \sim \frac{N\text{d}l}{R}
\text{.}
\end{equation} 
If we let the size of the polymer in the bulk phase, $R$, be proportional to $N^{\nu}$, then we can obtain $n_{\text{ads}}\sim N/N^{\nu} \sim N^{1-\nu}$. Recall that each adsorbed monomer occupies almost the same area on the surface in Fig. \ref{fig-layer} since the total number of monomers adsorbed on the surface, which corresponds to an area under the curve of the first peak, is regardless of $N$. From this, $n_{\text{ads}}$ is proportional to the area occupied by an adsorbed polymer, and $1/n_{\text{ads}}$ is proportional to the number of adsorbed polymers per unit area. The adsorbed amount, $\Gamma$, is the number of adsorbed polymers per unit area multiplied by the number of monomer in a polymer, so finally, we can obtain that $\Gamma\sim N^{\delta} \sim n_{\text{ads}}^{-1}\times N ~\sim N^{\nu}$, which reduces to $\nu=\delta$ for the ideal chain adsorbed on the reflecting surface. Discrepancy of two exponents at small $N$ can be also explained by the comparable lengths between the local chain non-ideality and the polymer size.

\begin{figure}[t]
	\centering
	\includegraphics[width=3.1in]{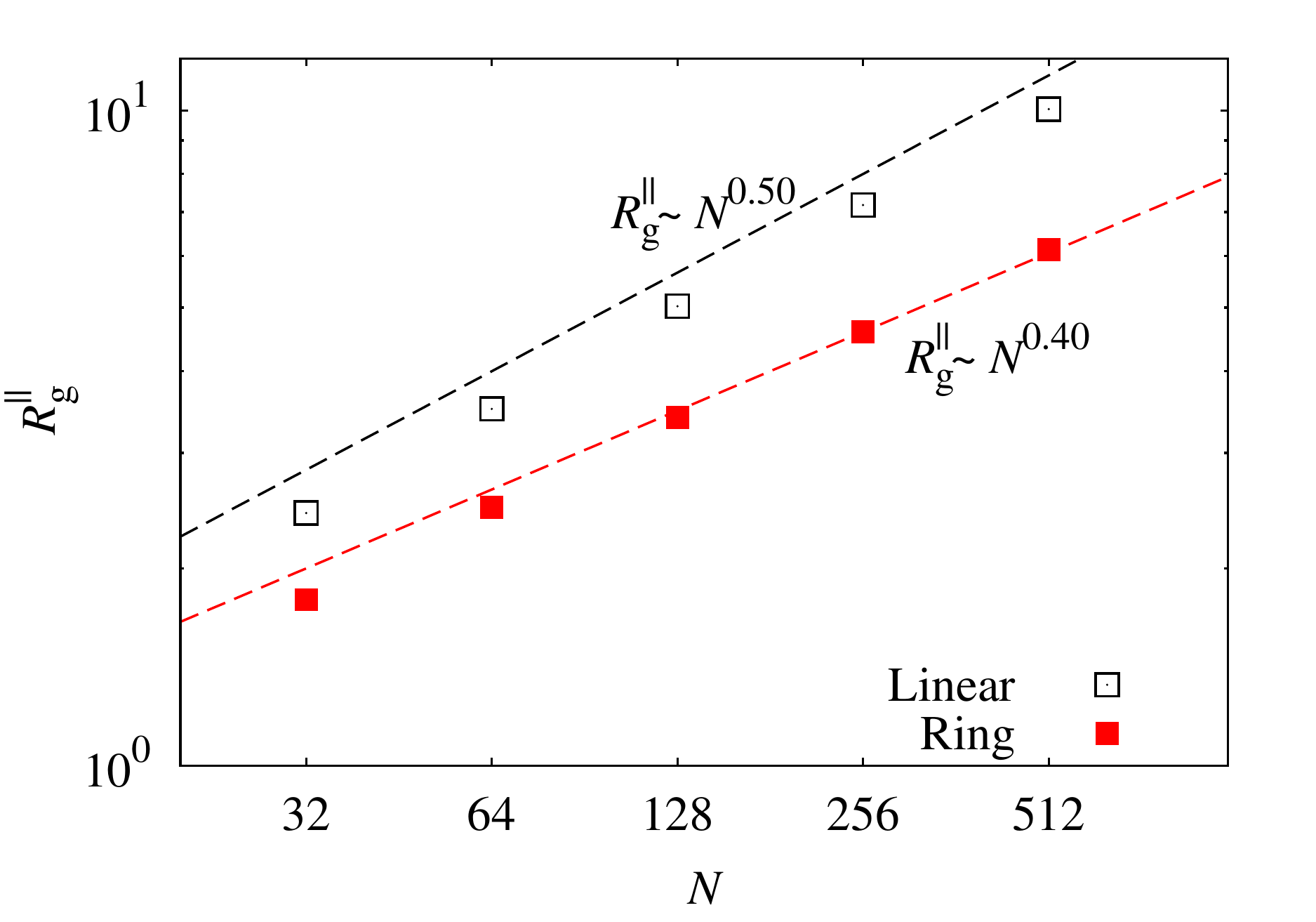} \caption{\label{fig-rgpar} Log-log plots of parallel components of gyration radii to the surface versus $N$ for linear (open squares) and ring (filled squares) polymers. Black and red dashed lines represent $R_{\text{g}}\sim N^{0.50}$ and $\sim N^{0.40}$, respectively.}
\end{figure}

It is an interesting problem to investigate why two exponents of ring polymers are similar to each other for large $N$ even though they are not ideal. In Eqn. \ref{eqn-idealads}, we used the size of polymers in bulk phase, $R$, when calculating the volume of the disk and self-monomer density because they are considered as ideal chains. However, rings swell along parallel direction to the surface as observed in Fig. \ref{fig-rgslab}, thus we can not directly substitute the size of the adsorbed polymer into its bulk size in Eqn. \ref{eqn-idealads}. Instead, the volume of disk $V_{\text{disk}}$, is proportional to $R_{\text{g}}^{||2}\text{d}l$ and the self-monomer density is proportional to $N/V_{\text{ring}}\approx N/R_{x}R_{y}R_{z}$ where $V_{\text{ring}}$ is the volume occupied by an adsorbed ring and $R_{\alpha}$ represents the gyration radius along $\alpha$-direction ($\alpha=x$, $y$, $z$). Thereby, $n_{\text{ads}}$ for the ring is:
\begin{equation}
\label{eqn-ringads}
n_{\text{ads}} \approx V_{\text{disk}}\times\frac{N}{V_{\text{ring}}}\approx R^{||2}\text{d}l\times \frac{N}{R_{x}R_{y}R_{z}} 
\text{.}
\end{equation} 
Figure \ref{fig-rgpar} shows log-log plots of parallel size of linear and ring polymers versus $N$. In this figure, $R_{\text{g}}^{||}$ seems to follow scaling behavior with exponent, $\nu '$, and fitted exponents for linear and ring polymers are 0.51 and 0.44, respectively. $\nu'$ of linear polymers is very close to the bulk one, $\nu=0.5$, which is supported by the Silberberg hypothesis. However, it should be noted that the size of adsorbed ring polymers, even though their parallel sizes increase compared to those in the bulk as in Fig. \ref{fig-rgslab}, are scaled by the exponent which is very similar to Flory exponent. Especially, for long rings of $N=256$ and $512$, parallel polymer sizes are proportional to $N^{2/5}$, which yields the same scaling behavior as the bulk ring size. As a result, the scaling exponent of $V_{\text{disk}}$ versus $N$ does not change while we replace $R_{\text{g}}^{||}$ by $R$ in Eqn. \ref{eqn-ringads}. 

It is also desirable to compare $R_{x,\text{bulk}}^3$ ($\approx R^3$ in Eqn. \ref{eqn-idealads}) with $R_x R_y R_z$ for the valid substitution of the bulk self-monomer density into the adsorbed one, where $R_{x,\text{bulk}}$ is radius of gyration along $x-$ direction of a bulk polymer.  $R_x R_y R_z$ and $R_{x,\text{bulk}}^3$ represent, approximately, $V_{\text{ring}}$ in the film and in bulk phase, respectively. In Tab. \ref{tab-polvol}, it is shown that the volume of a linear polymer adsorbed on the surface is much smaller than that in bulk phase because the parallel size is decoupled from the perpendicular one. The volume of an adsorbed ring polymer is, however, very similar with a bulk ring. This observation reflects the highly packed structure of a ring because reduction of perpendicular ring size is compensated by increase of parallel size. The better consistency between $R^3$ and $R_{x}R_{y}R_{z}$ is expected for the more compact ring polymers of $N\rightarrow \infty$. According to this argument, by substituting $R^3$ into $R_{x}R_{y}R_{z}$, Eqn. \ref{eqn-ringads} arrives at the same relation with the ideal chain (Eqn. \ref{eqn-idealads}) which reproduces $\nu\approx\delta$.

\begin{table*}[t]
\caption{\label{tab-polvol} Gyration radii in different directions and multiplication of three gyration radii in the film and bulk systems. Because gyration radii in different directions of a bulk polymer are almost the same, we use $R_{x,\text{bulk}}^3$ instead of $R_{x,\text{bulk}}R_{y,\text{bulk}}R_{z,\text{bulk}}$ in Eqn. \ref{eqn-idealads}.}
\setlength{\tabcolsep}{15pt}
\centering
\begin{tabular}{rrrrrrrrrr}
\hline\hline
 & \multicolumn{4}{c}{Linear} & &\multicolumn{4}{c}{Ring} \\
\cline{2-5}\cline{7-10}
$N$ &  $R_{x}$ & $R_{z}$ &$R_{x}R_{y}R_{z}$ & $R_{x,\text{bulk}}^{3}$ & 
&  $R_{x}$ & $R_{z}$ &$R_{x}R_{y}R_{z}$ & $R_{x,\text{bulk}}^{3}$  \\
\hline
32 & 1.71 & 1.48 & 4.38 & 4.52 & &1.27 & 1.10 & 1.76 & 1.78 \\
64 & 2.48 & 2.14 & 13.11 & 13.88 & & 1.75 & 1.53 & 4.71 & 4.78 \\
128 & 3.55 & 3.08 & 38.94 & 41.31 & & 2.40 & 2.11 & 12.22 & 12.35 \\
256 & 5.05 & 4.42 & 113.77 & 119.43 & & 3.25 & 2.86 & 30.10 & 30.85 \\
512 & 7.20 & 6.59 & 323.51 & 350.47 & & 4.34 & 3.81 & 71.76 & 73.09 \\
\hline\hline
\end{tabular}
\end{table*}

The exponent $\delta$ can be also obtained by the relation between the adsorbed amount and the train properties, using scaling exponents, $\gamma_{\text{n}}$ and $\gamma_{\text{s}}$. If there is no correlation between $n_{\text{tr}}$ and $s_{tr}$, the product of two averaged qantities, $\langle n_{\text{tr}}\rangle\times\langle s_{\text{tr}}\rangle$ indicates the average number of adsorbed monomers per an adsorbed polymer, $n_{\text{ads}}$. Because $1/n_{\text{ads}}$ is proportional to the number of adsorbed polymers on unit area and $\Gamma /N \sim N^{\delta -1}$ as mentioned above, the equation, $1/n_{\text{ads}}\sim N^{-\gamma_{\text{n}}-\gamma_{\text{s}}}\sim N^{\delta-1}$, arrives at $\delta=1-\gamma_{\text{n}}-\gamma_{\text{s}}$. Fitted scaling exponents of $\gamma_{\text n}$ and $\gamma_{\text s}$ with $\delta$ are listed in Table \ref{tab-adsexp}. Because the summation of three exponents equals unity, the number of trains and the size of train are uncorrelated. The relation, $\delta=1-\gamma_{\text{n}}-\gamma_{\text{s}}$, reflects that the dependence of $N$ on the adsorbed amount is determined by those of train size and the number. $\gamma_{\text{s}}$ for linear and ring polymers are almost the same with each other and are close to zero, which means that the effect of train size on the change of adsorbed amount is negligible. The main factor determining different adsorbed amounts between linear and ring polymers is the number of trains, which, in turn, is due to the difference in self-monomer densities.

\subsection{Coordination Number of Polymers}

\begin{figure}[t]
	\centering
	\includegraphics[width=3.1in]{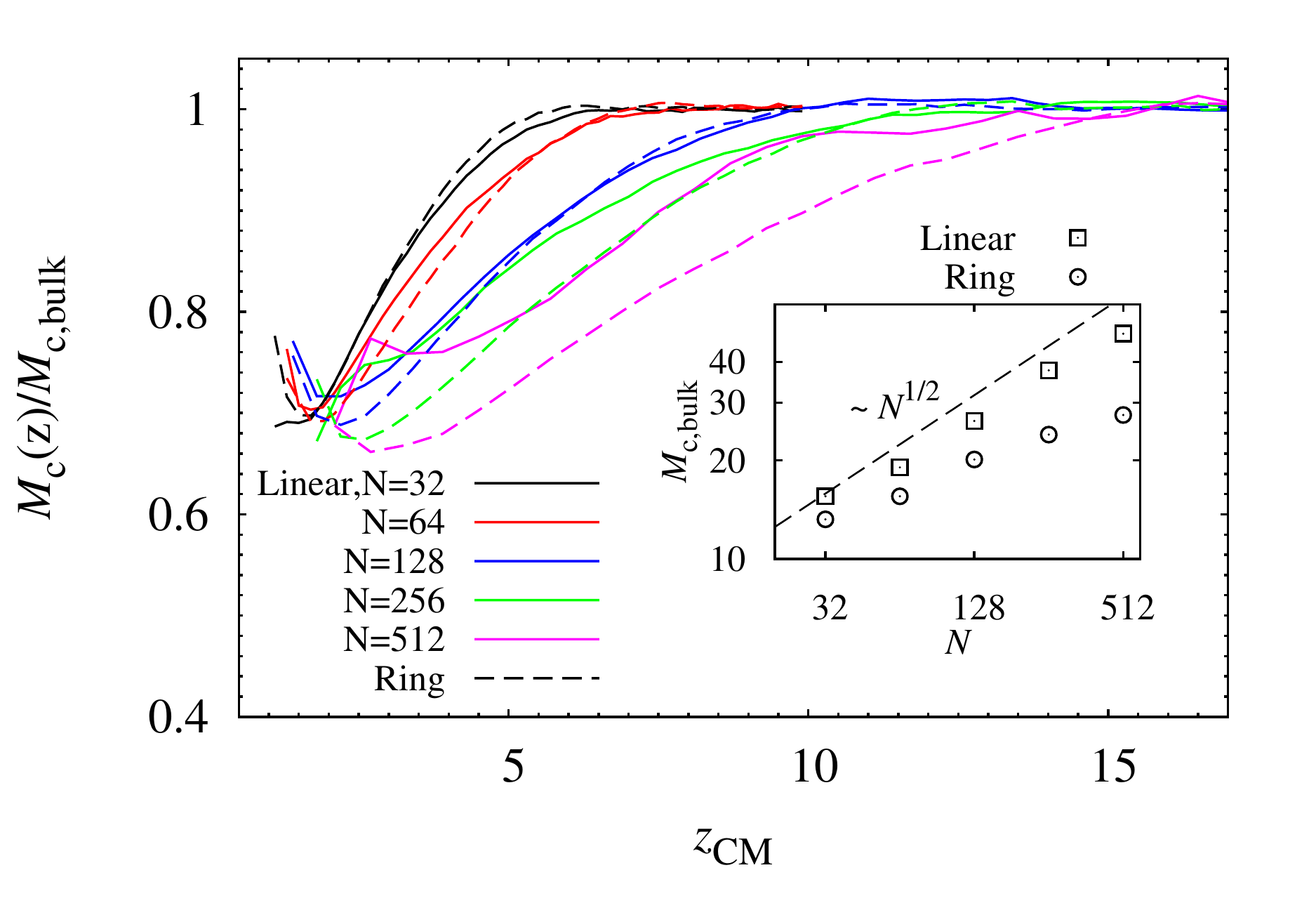} 
	\caption{\label{fig-cnvsz} Coordination numbers normalized by bulk values for linear (solid lines) and ring (dashed lines) polymers as a function of the distance of the polymer center-of-mass from the surface. Different colors represent different lengths of polymers. An inset shows the coordination numbers of linear and ring polymers in bulk phases. }
\end{figure}

Another striking result in static properties of confined polymers is the decrease of the coordination number of a polymer, which is a good measure of conformational properties of polymeric melts. It is well known that while the coordination number of the linear polymer grows with $N^{1/2}$, that of the ring polymer saturates at $N\rightarrow \infty$ (inset of Fig. \ref{fig-cnvsz}).\cite{Halverson_JChemPhys_2011}  We evaluate the coordination numbers normalized by their bulk values as a function of $z_{\text CM}$ in Fig. \ref{fig-cnvsz}. Here, the number of contacting molecular pairs per a polymer is defined by the coordination number, where two polymers of which at least one monomer-monomer pair is closer in space than $\sigma$ are regarded as the contacting pair. Note that the amount of decrease for the ring is larger than that for the linear counterpart except for the $N=32$ case. This observation reveals that the repulsive topological interaction due to the reflecting blobs makes the locally tangled neighboring rings squeeze out. From this result, it is strongly suspected that the confinement effect contributes to forming a segregated chromosome in a nucleus.  For $N=32$ linear case, the small chain size comparable to the non-ideality in the vicinity of the surface leads to the exceptional drop of the number of surrounding molecules.

\section{\label{sec-4} Conclusions}

In this work, we have presented structures of confined ring polymers in a melt by comparing those of ideal chains. Ring polymers in a melt are known to form compact structure by packing ideal blobs. The length scale of local structure ordering due to the surface-monomer excluded volume interaction is shorter than the size of the ideal blob, which leads to the very similar local structures between linear and ring polymers. According to the Silberberg's reflecting boundary argument, the reflecting blobs of linear polymers do not feel the excluded volume interaction. In contrast, reflecting parts of ring polymers feel the repulsion with other blobs due to the topological excluded volume interactions. As a result, the confined ring crowds out the neighboring molecules which makes itself more compact, and it expands along parallel direction to the surface. Because we have yet to reach very long rings in asymptotic limit, the exponents obtained in this work, $\nu$, $\gamma$, $\delta_{\text n}$, and $\delta_{\text s}$ are not universal. Nevertheless, we found interesting relations between these exponents which are expected to also hold for very long rings due to the stronger topological effect than our rings. Moreover, longer rings are expected to show the larger amount of parallel size expansion and a bigger drop of the coordination number than our ring polymers. Our observations provide the possibility that segregated conformation of chromosomes is originated from the confinement effect in a nucleus envelope in addition to the factors mentioned in the introduction, e.g., a long disentanglement time of a chromatin fiber and a telomeric region at the end. In this view point, it is challenging to investigate variation of crossover lengths of the ring polymer structure from ideal to globular statistics under the confinement. In addition to the confinement effect, various aspects for the interplay between the physics of ring polymer melt and the chromosome packing are currently under study.


\section*{Acknowledgements}

We would like to thank Kurt Kremer for the discussions in the early stage of this study. We also acknowledge the financial support from the Korean National Research Foundation (Grant Nos. NRF-2012R1A1A2042062 and NRF-2007-0056095). This work is also supported by the IRTG 1404 program (No. 2011-0032203) of National Research Foundation of Korea (NRF) and the Brain Korea (BK) 21 Plus program. Computational resources have been provided by KISTI supercomputing center through Grand No. KSC-2013-C2-021.




\bibliography{paper} 

\end{document}